\documentclass[11pt, a4paper]{article}
\PassOptionsToPackage{unicode}{hyperref}
\usepackage{jheppub}
\usepackage{upgreek}
\usepackage{slashed}

\title{Chiral algebras from $\Omega$-deformation}

\author[1,2]{Jihwan Oh}
\author[1]{and Junya Yagi}

\affiliation[1]{Perimeter Institute for Theoretical Physics,
  Waterloo, ON
  N2L 2Y5 Canada}
\affiliation[2]{Department of Physics,
  University of California, Berkeley, CA 94720 USA}

\abstract{In the presence of an $\Omega$-deformation, local operators
  generate a chiral algebra in the topological--holomorphic twist of a
  four-dimensional $\CN = 2$ supersymmetric field theory.  We show
  that for a unitary $\CN = 2$ superconformal field theory, the chiral
  algebra thus defined is isomorphic to the one introduced by Beem et
  al.  Our definition of the chiral algebra covers nonconformal
  theories with insertions of suitable surface defects.}


\arxivnumber{}

\let\U\relax
\let\C\relax

\newcommand{\gf}{\mathfrak{g}}

\newcommand{\del}{\partial}
\newcommand{\delb}{{\bar\partial}}

\newcommand{\vol}{\mathrm{vol}}

\renewcommand{\Re}{\mathop{\mathrm{Re}}\nolimits}

\newcommand{\Tr}{\mathop{\mathrm{Tr}}\nolimits}

\newcommand{\SU}{\mathrm{SU}}

\newcommand{\GL}{\mathrm{GL}}

\newcommand{\U}{\mathrm{U}}

\newcommand{\iso}{\cong}

\newcommand{\Z}{\mathbb{Z}}

\newcommand{\R}{\mathbb{R}}
\newcommand{\C}{\mathbb{C}}

\let\nc\newcommand
\let\renc\renewcommand
\nc{\wbar}{\overline}
\let\td\tilde
\let\wtd\widetilde
\let\wht\widehat
\let\mcl\mathcal

\nc{\ab}{{\bar{a}}} \nc{\at}{\tilde{a}} \nc{\ah}{\hat{a}}
\nc{\bb}{{\bar{b}}} \nc{\bt}{\tilde{b}} \nc{\bh}{\hat{b}}
\nc{\cb}{{\bar{c}}} \nc{\ct}{\tilde{c}} %\nc{\ch}{\hat{c}}
\nc{\db}{{\bar{d}}} \nc{\dt}{\tilde{d}} \renc{\dh}{\hat{d}}
\nc{\eb}{{\bar{e}}} \nc{\et}{\tilde{e}} \nc{\eh}{\hat{e}}
\nc{\fb}{{\bar{f}}} \nc{\ft}{\tilde{f}} \nc{\fh}{\hat{f}}
\nc{\gb}{{\bar{g}}} \nc{\gt}{\tilde{g}} \nc{\gh}{\hat{g}}
\nc{\hb}{{\bar{h}}} \nc{\hh}{\hat{h}} %\nc{\ht}{\tilde{h}}
\nc{\ib}{{\bar{\imath}}} \nc{\ih}{\hat{\imath}} %\nc{\it}{\tilde{\imath}}
\nc{\jb}{{\bar{\jmath}}} \nc{\jt}{\tilde{\jmath}} \nc{\jh}{\hat{\jmath}}
\nc{\kb}{{\bar{k}}} \nc{\kt}{\tilde{k}} \nc{\kh}{\hat{k}}
\nc{\lb}{{\bar{l}}} \nc{\lt}{\tilde{l}} \nc{\lh}{\hat{l}}
\nc{\mb}{{\bar{m}}} \nc{\mt}{\tilde{m}} \nc{\mh}{\hat{m}}
\nc{\nb}{{\bar{n}}} \nc{\nt}{\tilde{n}} \nc{\nh}{\hat{n}}
\nc{\ob}{{\bar{o}}} \nc{\ot}{\tilde{o}} \nc{\oh}{\hat{o}}
\nc{\pb}{{\bar{p}}} \nc{\pt}{\tilde{p}} \nc{\ph}{\hat{p}}
\nc{\qb}{{\bar{q}}} \nc{\qt}{\tilde{q}} \nc{\qh}{\hat{q}}
\nc{\rb}{{\bar{r}}} \nc{\rt}{\tilde{r}} \nc{\rh}{{\hat{r}}}
\renc{\sb}{{\bar{s}}} \nc{\st}{\tilde{s}} \nc{\sh}{\hat{s}}
\nc{\tb}{{\bar{t}}} \renc{\th}{\hat{t}} %\nc{\tt}{\tilde{t}}
\nc{\ub}{{\bar{u}}} \nc{\ut}{\tilde{u}} \nc{\uh}{\hat{u}}
\nc{\vb}{{\bar{v}}} \nc{\vt}{\tilde{v}} \nc{\vh}{\hat{v}}
\nc{\wb}{{\bar{w}}} \nc{\wt}{\tilde{w}} \nc{\wh}{\hat{w}}
\nc{\xb}{{\bar{x}}} \nc{\xt}{\tilde{x}} \nc{\xh}{\hat{x}}
\nc{\yb}{{\bar{y}}} \nc{\yt}{\tilde{y}} \nc{\yh}{\hat{y}}
\nc{\zb}{{\bar{z}}} \nc{\zt}{\tilde{z}} \nc{\zh}{\hat{z}}

\nc{\Ab}{{\wbar{A}}} \nc{\At}{{\wtd{A}}} \nc{\Ah}{{\wht{A}}}
\nc{\Bb}{{\wbar{B}}} \nc{\Bt}{{\wtd{B}}} \nc{\Bh}{{\wht{B}}}
\nc{\Cb}{{\wbar{C}}} \nc{\Ct}{{\wtd{C}}} \nc{\Ch}{{\wht{C}}}
\nc{\Db}{{\wbar{D}}} \nc{\Dt}{{\wtd{D}}} \nc{\Dh}{{\wht{D}}}
\nc{\Eb}{{\wbar{E}}} \nc{\Et}{{\wtd{E}}} \nc{\Eh}{{\wht{E}}}
\nc{\Fb}{{\wbar{F}}} \nc{\Ft}{{\wtd{F}}} \nc{\Fh}{{\wht{F}}}
\nc{\Gb}{{\wbar{G}}} \nc{\Gt}{{\wtd{G}}} \nc{\Gh}{{\wht{G}}}
\nc{\Hb}{{\wbar{H}}} \nc{\Ht}{{\wtd{H}}} \nc{\Hh}{{\wht{H}}}
\nc{\Ib}{{\bar{I}}} \nc{\It}{{\wtd{I}}} \nc{\Ih}{{\wht{I}}}
\nc{\Jb}{{\wbar{J}}} \nc{\Jt}{{\wtd{J}}} \nc{\Jh}{{\wht{J}}}
\nc{\Kb}{{\wbar{K}}} \nc{\Kt}{{\wtd{K}}} \nc{\Kh}{{\wht{K}}}
\nc{\Lb}{{\wbar{L}}} \nc{\Lt}{{\wtd{L}}} \nc{\Lh}{{\wht{L}}}
\nc{\Mb}{{\wbar{M}}} \nc{\Mt}{{\wtd{M}}} \nc{\Mh}{{\wht{M}}}
\nc{\Nb}{{\wbar{N}}} \nc{\Nt}{{\wtd{N}}} \nc{\Nh}{{\wht{N}}}
\nc{\Ob}{{\wbar{O}}} \nc{\Ot}{{\wtd{O}}} \nc{\Oh}{{\wht{O}}}
\nc{\Pb}{{\wbar{P}}} \nc{\Pt}{{\wtd{P}}} \nc{\Ph}{{\wht{P}}}
\nc{\Qb}{{\wbar{Q}}} \nc{\Qt}{{\wtd{Q}}} \nc{\Qh}{{\wht{Q}}}
\nc{\Rb}{{\wbar{R}}} \nc{\Rt}{{\wtd{R}}} \nc{\Rh}{{\wht{R}}}
\nc{\Sb}{{\wbar{S}}} \nc{\St}{{\wtd{S}}} \nc{\Sh}{{\wht{S}}}
\nc{\Tb}{{\wbar{T}}} \nc{\Tt}{{\wtd{T}}} \nc{\Th}{{\wht{T}}}
\nc{\Ub}{{\wbar{U}}} \nc{\Ut}{{\wtd{U}}} \nc{\Uh}{{\wht{U}}}
\nc{\Vb}{{\wbar{V}}} \nc{\Vt}{{\wtd{V}}} \nc{\Vh}{{\wht{V}}}
\nc{\Wb}{{\wbar{W}}} \nc{\Wt}{{\wtd{W}}} \nc{\Wh}{{\wht{W}}}
\nc{\Xb}{{\wbar{X}}} \nc{\Xt}{{\wtd{X}}} \nc{\Xh}{{\wht{X}}}
\nc{\Yb}{{\wbar{Y}}} \nc{\Yt}{{\wtd{Y}}} \nc{\Yh}{{\wht{Y}}}
\nc{\Zb}{{\wbar{Z}}} \nc{\Zt}{{\wtd{Z}}} \nc{\Zh}{{\wht{Z}}}

\nc{\CA}{{\mcl{A}}} \nc{\CAb}{{\wbar{\CA}}} \nc{\CAt}{{\wtd{\CA}}} \nc{\CAh}{{\wht{\CA}}}
\nc{\CB}{{\mcl{B}}} \nc{\CBb}{{\wbar{\CB}}} \nc{\CBt}{{\wtd{\CB}}} \nc{\CBh}{{\wht{\CB}}}
\nc{\CC}{{\mcl{C}}} \nc{\CCb}{{\wbar{\CC}}} \nc{\CCt}{{\wtd{\CC}}} \nc{\CCh}{{\wht{\CC}}}
\nc{\cD}{{\mcl{D}}} \nc{\cDb}{{\wbar{\cD}}} \nc{\cDt}{{\wtd{\cC}}} \nc{\cDh}{{\wht{\cD}}}
\nc{\CE}{{\mcl{E}}} \nc{\CEb}{{\wbar{\CE}}} \nc{\CEt}{{\wtd{\CE}}} \nc{\CEh}{{\wht{\CE}}}
\nc{\CF}{{\mcl{F}}} \nc{\CFb}{{\wbar{\CF}}} \nc{\CFt}{{\wtd{\CF}}} \nc{\CFh}{{\wht{\CF}}}
\nc{\CG}{{\mcl{G}}} \nc{\CGb}{{\wbar{\CG}}} \nc{\CGt}{{\wtd{\CG}}} \nc{\CGh}{{\wht{\CG}}}
\nc{\CH}{{\mcl{H}}} \nc{\CHb}{{\wbar{\CH}}} \nc{\CHt}{{\wtd{\CH}}} \nc{\CHh}{{\wht{\CH}}}
\nc{\CI}{{\mcl{I}}} \nc{\CIb}{{\wbar{\CI}}} \nc{\CIt}{{\wtd{\CI}}} \nc{\CIh}{{\wht{\CI}}}
\nc{\CJ}{{\mcl{J}}} \nc{\CJb}{{\wbar{\CJ}}} \nc{\CJt}{{\wtd{\CJ}}} \nc{\CJh}{{\wht{\CJ}}}
\nc{\CK}{{\mcl{K}}} \nc{\CKb}{{\wbar{\CK}}} \nc{\CKt}{{\wtd{\CK}}} \nc{\CKh}{{\wht{\CK}}}
\nc{\CL}{{\mcl{L}}} \nc{\CLb}{{\wbar{\CL}}} \nc{\CLt}{{\wtd{\CL}}} \nc{\CLh}{{\wht{\CL}}}
\nc{\CM}{{\mcl{M}}} \nc{\CMb}{{\wbar{\CM}}} \nc{\CMt}{{\wtd{\CM}}} \nc{\CMh}{{\wht{\CM}}}
\nc{\CN}{{\mcl{N}}} \nc{\CNb}{{\wbar{\CN}}} \nc{\CNt}{{\wtd{\CN}}} \nc{\CNh}{{\wht{\CN}}}
\nc{\CO}{{\mcl{O}}} \nc{\COb}{{\wbar{\CO}}} \nc{\COt}{{\wtd{\CO}}} \nc{\COh}{{\wht{\CO}}}
\nc{\CP}{{\mcl{P}}} \nc{\CPb}{{\wbar{\CP}}} \nc{\CPt}{{\wtd{\CP}}} \nc{\CPh}{{\wht{\CP}}}
\nc{\CQ}{{\mcl{Q}}} \nc{\CQb}{{\wbar{\CQ}}} \nc{\CQt}{{\wtd{\CQ}}} \nc{\CQh}{{\wht{\CQ}}}
\nc{\CR}{{\mcl{R}}} \nc{\CRb}{{\wbar{\CR}}} \nc{\CRt}{{\wtd{\CR}}} \nc{\CRh}{{\wht{\CR}}}
\nc{\CS}{{\mcl{S}}} \nc{\CSb}{{\wbar{\CS}}} \nc{\CSt}{{\wtd{\CS}}} \nc{\CSh}{{\wht{\CS}}}
\nc{\CT}{{\mcl{T}}} \nc{\CTb}{{\wbar{\CT}}} \nc{\CTt}{{\wtd{\CT}}} \nc{\CTh}{{\wht{\CT}}}
\nc{\CU}{{\mcl{U}}} \nc{\CUb}{{\wbar{\CU}}} \nc{\CUt}{{\wtd{\CU}}} \nc{\CUh}{{\wht{\CU}}}
\nc{\CV}{{\mcl{V}}} \nc{\CVb}{{\wbar{\CV}}} \nc{\CVt}{{\wtd{\CV}}} \nc{\CVh}{{\wht{\CV}}}
\nc{\CW}{{\mcl{W}}} \nc{\CWb}{{\wbar{\CW}}} \nc{\CWt}{{\wtd{\CW}}} \nc{\CWh}{{\wht{\CW}}}
\nc{\CX}{{\mcl{X}}} \nc{\CXb}{{\wbar{\CX}}} \nc{\CXt}{{\wtd{\CX}}} \nc{\CXh}{{\wht{\CX}}}
\nc{\CY}{{\mcl{Y}}} \nc{\CYb}{{\wbar{\CY}}} \nc{\CYt}{{\wtd{\CY}}} \nc{\CYh}{{\wht{\CY}}}
\nc{\CZ}{{\mcl{Z}}} \nc{\CZb}{{\wbar{\CZ}}} \nc{\CZt}{{\wtd{\CZ}}} \nc{\CZh}{{\wht{\CZ}}}

\let\eps\epsilon
\let\ups\upsilon
\let\veps\varepsilon
\let\vtht\vartheta

\let\vsgm\varsigma
\let\vphi\varphi
\let\vrho\varrho

\nc{\alphab}{{\bar{\alpha}}} \nc{\alphat}{{\td{\alpha}}} \nc{\alphah}{{\hat{\alpha}}}
\nc{\betab}{{\bar{\beta}}}   \nc{\betat}{{\td{\beta}}}   \nc{\betah}{{\hat{\beta}}} 
\nc{\gammab}{{\bar{\gamma}}} \nc{\gammat}{{\td{\gamma}}} \nc{\gammah}{{\hat{\gamma}}} 
\nc{\deltab}{{\bar{\delta}}} \nc{\deltat}{{\td{\delta}}} \nc{\deltah}{{\hat{\delta}}} 
\nc{\epsilonb}{{\bar{\eps}}} \nc{\epsilont}{{\td{\eps}}} \nc{\epsilonh}{{\hat{\eps}}} 
\nc{\vepsb}{{\bar{\veps}}}   \nc{\vepst}{{\td{\veps}}}   \nc{\vepsh}{{\hat{\veps}}} 
\nc{\zetab}{{\bar{\zeta}}}   \nc{\zetat}{{\td{\zeta}}}   \nc{\zetah}{{\hat{\zeta}}} 
\nc{\etab}{{\bar{\eta}}}     \nc{\etat}{{\td{\eta}}}     \nc{\etah}{{\hat{\eta}}} 
\nc{\thetab}{{\bar{\theta}}} \nc{\thetat}{{\td{\theta}}} \nc{\thetah}{{\hat{\theta}}} 
\nc{\vthetab}{{\bar{\vtht}}} \nc{\vthetat}{{\td{\vtht}}} \nc{\vthetah}{{\hat{\vtht}}} 
\nc{\lambdab}{{\bar{\lambda}}} \nc{\lambdat}{{\td{\lambda}}} \nc{\lambdah}{{\hat{\lambda}}} 
\nc{\iotab}{{\bar{\iota}}}   \nc{\iotat}{{\td{\iota}}}   \nc{\iotah}{{\hat{\iota}}} 
\nc{\kappab}{{\bar{\kappa}}} \nc{\kappat}{{\td{\kappa}}} \nc{\kappah}{{\hat{\kappa}}} 
\nc{\lmdb}{{\bar{\lmd}}}     \nc{\lmdt}{{\td{\lmd}}}     \nc{\lmdh}{{\hat{\lmd}}} 
\nc{\mub}{{\bar{\mu}}}       \nc{\mut}{{\td{\mu}}}       \nc{\muh}{{\hat{\mu}}} 
\nc{\nub}{{\bar{\nu}}}       \nc{\nut}{{\td{\nu}}}       \nc{\nuh}{{\hat{\nu}}} 
\nc{\xib}{{\bar{\xi}}}       \nc{\xit}{{\td{\xi}}}       \nc{\xih}{{\hat{\xi}}} 
\nc{\pib}{{\bar{\pi}}}       \nc{\pit}{{\td{\pi}}}       \nc{\pih}{{\hat{\pi}}} 
\nc{\vpib}{{\bar{\vpi}}}     \nc{\vpit}{{\td{\vpi}}}     \nc{\vpih}{{\hat{\vpi}}} 
\nc{\rhob}{{\bar{\rho}}}     \nc{\rhot}{{\td{\rho}}}     \nc{\rhoh}{{\hat{\rho}}} 
\nc{\vrhob}{{\bar{\vrho}}}   \nc{\vrhot}{{\td{\vrho}}}   \nc{\vrhoh}{{\hat{\vrho}}} 
\nc{\sigmab}{{\bar{\sigma}}} \nc{\sigmat}{{\td{\sigma}}} \nc{\sigmah}{{\hat{\sigma}}} 
\nc{\vsigmab}{{\bar{\vsgm}}} \nc{\vsigmat}{{\td{\vsgm}}} \nc{\vsigmah}{{\hat{\vsgm}}} 
\nc{\taub}{{\bar{\tau}}}     \nc{\taut}{{\td{\tau}}}     \nc{\tauh}{{\hat{\tau}}} 
\nc{\upsb}{{\bar{\ups}}} \nc{\upst}{{\td{\ups}}} \nc{\upsh}{{\hat{\ups}}} 
\nc{\phib}{{\bar{\phi}}}     \nc{\phit}{{\td{\phi}}}     \nc{\phih}{{\hat{\phi}}} 
\nc{\varphib}{{\bar{\vphi}}}   \nc{\varphit}{{\td{\vphi}}}   \nc{\varphih}{{\hat{\vphi}}} 
\nc{\chib}{{\bar{\chi}}}     \nc{\chit}{{\td{\chi}}}     \nc{\chih}{{\hat{\chi}}} 
\nc{\psib}{{\bar{\psi}}}     \nc{\psit}{{\td{\psi}}}     \nc{\psih}{{\hat{\psi}}} 
\nc{\omegab}{{\bar{\omega}}} \nc{\omegat}{{\td{\omega}}} \nc{\omegah}{{\hat{\omega}}} 

\nc{\Gammab}{{\wbar{\Gamma}}}     \nc{\Gammat}{{\wtd{\Gamma}}}     \nc{\Gammah}{{\wht{\Gamma}}}
\nc{\Deltab}{{\wbar{\Delta}}}     \nc{\Deltat}{{\wtd{\Delta}}}     \nc{\Deltah}{{\wht{\Delta}}}
\nc{\Thetab}{{\wbar{\Theta}}}     \nc{\Thetat}{{\wtd{\Theta}}}     \nc{\Thetah}{{\wht{\Theta}}}
\nc{\Lambdab}{{\wbar{\Lambda}}}   \nc{\Lambdat}{{\wtd{\Lambda}}}   \nc{\Lambdah}{{\wht{\Lambda}}}
\nc{\Xib}{{\wbar{\Xi}}}           \nc{\Xit}{{\wtd{\Xi}}}           \nc{\Xih}{{\wht{\Xi}}}
\nc{\Pib}{{\wbar{\Pi}}}           \nc{\Pit}{{\wtd{\Pi}}}           \nc{\Pih}{{\wht{\Pi}}}
\nc{\Sigmab}{{\wbar{\Sigma}}}     \nc{\Sigmat}{{\wtd{\Sigma}}}     \nc{\Sigmah}{{\wht{\Sigma}}}
\nc{\Upsilonb}{{\wbar{\Upsilon}}} \nc{\Upsilont}{{\wtd{\Upsilon}}} \nc{\Upsilonh}{{\wht{\Upsilon}}}
\nc{\Phib}{{\wbar{\Phi}}} \nc{\Phit}{{\wtd{\Phi}}} \nc{\Phih}{{\wht{\Phi}}}
\nc{\Psib}{{\wbar{\Psi}}}         \nc{\Psit}{{\wtd{\Psi}}}         \nc{\Psih}{{\wht{\Psi}}}
\nc{\Omegab}{{\wbar{\Omega}}}     \nc{\Omegat}{{\wtd{\Omega}}}     \nc{\Omegah}{{\wht{\Omega}}}

\newcommand{\rmd}{\mathrm{d}}

\newcommand{\ChS}{\mathrm{CS}}

\newcommand{\auxF}{\mathsf{F}}
\newcommand{\auxFb}{\overline{\auxF}}
\newcommand{\auxD}{\mathsf{D}}
\newcommand{\auxB}{\mathsf{B}}

\newcommand{\iu}{\mathrm{i}}

\let\starx\star
\let\star\relax
\newcommand{\star}{\mathop{\starx}\nolimits}

\newcommand{\SV}{S_{\text{V}}}

\newcommand{\BRST}{\text{BRST}}

\usepackage[mathscr]{euscript}

\newcommand{\SG}{\mathscr{G}}
\newcommand{\SR}{\mathscr{R}}
\renewcommand{\SV}{\mathscr{V}}

\newcommand{\Mod}{\mathscr{M}}
\newcommand{\Lag}{\mathscr{L}}

\newcommand{\Qone}{\mathbf{Q}}
\newcommand{\QQ}{\smash{\slashed{Q}}}
\newcommand{\QQt}{\smash{\widetilde{\slashed{Q}}}}
\newcommand{\QK}{Q_{\mathrm{K}}}
\newcommand{\QQK}{\QQ_{\mathrm{K}}}
\newcommand{\QQtK}{\QQt_{\mathrm{K}}}

\newcommand{\RW}{{\mathrm{RW}}}
\newcommand{\mRW}{{\mathrm{mRW}}}
\newcommand{\adj}{\mathrm{adj}}

\newcommand{\betaSD}{\upbeta}
\newcommand{\gammaSD}{\upgamma}
\newcommand{\bSD}{\mathtt{b}}
\newcommand{\cSD}{\mathtt{c}}

\begin{document}
\maketitle

\section{Introduction}

A remarkable property of any $\CN = 2$ superconformal field theory
(SCFT) in four dimensions is that it possesses an infinite-dimensional
symmetry in a certain protected sector.  This symmetry is generated by
local operators which depend holomorphically on a plane, just as the
chiral algebra of a two-dimensional conformal field theory is
generated by holomorphic currents.  Since their discovery by Beem et
al.~\cite{Beem:2013sza}, the chiral algebras of $\CN = 2$ SCFTs have
been studied intensely and shed new light on the physics of these
theories, such as unitarity bounds on central
charges~\cite{Beem:2013sza, Liendo:2015ofa, Lemos:2015orc}.

Prior to the introduction of the chiral algebras
in~\cite{Beem:2013sza}, Kapustin~\cite{Kapustin:2006hi} had come up
with another way to extract the structures of two-dimensional
holomorphic field theories from $\CN = 2$ supersymmetric field
theories.  Known as the topological--holomorphic twist, it is a
twist~\cite{Witten:1988ze, Witten:1988xj} of theories placed on a
product $\Sigma \times C$ of two surfaces.  As the name suggests, it
renders the theories topological on $\Sigma$ and holomorphic on $C$.

It has been suggested by Kevin Costello (and mentioned
in~\cite{Costello:2018fnz}) that an
$\Omega$-deformation~\cite{Nekrasov:2002qd, Nekrasov:2003rj} of
Kapustin's construction should give rise to a chiral algebra.
Moreover, this chiral algebra has been suspected to be isomorphic to
the one introduced in~\cite{Beem:2013sza}; for standard gauge
theories, it has been shown by Dylan Butson~\cite{Butson:2019} that
this is indeed the case.  One of the primary purposes of this paper is
to establish this isomorphism for all unitary $\CN = 2$ SCFTs.

To this end we start in two dimensions.  In section~\ref{sec:2d}, we
formulate $\Omega$-deformations of the topological twists of
two-dimensional $\CN = (2,2)$ supersymmetric field theories.  It turns
out that an $\Omega$-deformation does not really deform a twisted
theory on $\R^2$ if the theory before the twist is not only
supersymmetric but also conformal: the only effect is that the
supercharge $Q$ that one uses to define cohomology is replaced with a
linear combination $Q^\hbar$ of Poincar\'e and conformal supercharges.
Furthermore, for a unitary $\CN = (2,2)$ SCFT, we show that the
$Q^\hbar$-cohomology of local operators is isomorphic to the
cohomology of local operators with respect to another supercharge
$\QQ^\hbar$ which squares to zero.  We also recall a localization
formula for $\Omega$-deformed B-twisted gauge theories constructed
from vector and chiral multiplets.

All of these results can be applied more or less directly to $\CN = 2$
supersymmetric field theories in four dimensions.  This is what we do
in section~\ref{sec:4d}.  The point is that the $\CN = 2$
superconformal algebra contains the global $\CN = (2,2)$
superconformal algebra as a subalgebra.  Correspondingly, Kapustin's
topological--holomorphic twist of an $\CN = 2$ supersymmetric field
theory on $\Sigma \times C$ may be regarded as the B-twist of an
$\CN = (2,2)$ supersymmetric field theory on $\Sigma$.  This B-twisted
theory can be subjected to an $\Omega$-deformation, and for
$\Sigma = \R^2$, the $Q^\hbar$-cohomology of local operators is
naturally a chiral algebra on $C$.

If the theory is a unitary $\CN = 2$ SCFT on $\R^2 \times \C$, this
chiral algebra may be described alternatively as the
$\QQ^\hbar$-cohomology of local operators.  This alternative
description turns out to coincide with the definition given
in~\cite{Beem:2013sza}.  The isomorphism in question is thus
established.

For an $\CN = 2$ superconformal gauge theory consisting of vector
multiplets and hypermultiplets, the localization formula immediately
tells us that the associated chiral algebra is a gauged $\beta\gamma$
system.  In fact, the path integral for the four-dimensional theory
localizes to that for the gauged $\beta\gamma$ system, placed directly
on the holomorphic surface $C$.  This is an advantage of our approach.

Another advantage is that we can define chiral algebras for
\emph{nonconformal} theories, provided that suitable surface defects
are inserted on $C$.  For our definition of the chiral algebra to
work, we only need to ensure that the $\CN = 2$ supersymmetric field
theory under consideration has a well-defined twisted rotation
symmetry on $\R^2$ and an $\Omega$-deformation can be turned on with
respect to it.  This is guaranteed if the R-symmetry used in the twist
along $\Sigma$ is nonanomalous -- a condition which is tied to
conformal invariance by $\CN = 2$ supersymmetry.  However, even if the
R-symmetry is anomalous, we can add an unusual $\theta$-term and make
the theory invariant under the twisted rotations.  The added term
breaks the gauge symmetry, but it can be restored by a gauge anomaly
from a surface defect which, in general, preserves only $\CN = (0,2)$
supersymmetry on $C$.  In this way we can obtain, for example, all
gauged $\beta\gamma$--$bc$ systems from gauge theories with surface
defects.

We emphasize that although for specific examples we consider theories
with Lagrangian descriptions, the existence of a Lagrangian is not
essential to our construction.  What is essential is realization of a
sensible $\Omega$-deformation on $\R^2$.  For $\CN = 2$ SCFTs, there
is a canonical definition whether or not a Lagrangian formulation
exists, and this definition, combined with unitarity, yields the
chiral algebras of~\cite{Beem:2013sza}.

Lastly, our approach applies equally well to $\CN = 4$ supersymmetric
field theories in three dimensions.  As we show in
section~\ref{sec:3d}, considerations similar to those described above
lead to the conclusion that an $\Omega$-deformation of the
Rozansky--Witten twist~\cite{Rozansky:1996bq} of an $\CN = 4$
supersymmetric gauge theory on $\R^2 \times \R$, constructed from
vector multiplets and hypermultiplets, is equivalent to a topological
gauged quantum mechanics.  For $\CN = 4$ SCFTs, we reproduce in this
manner some of the results obtained in~\cite{Beem:2016cbd,
  Dedushenko:2016jxl}.  Our approach clarifies the relation between
the $\QQ^\hbar$-cohomology, studied in~\cite{Beem:2016cbd}, and
deformation quantizations of the chiral rings.

In appendix \ref{sec:QV-vector-chiral}, we describe the supersymmetry
transformation laws for vector and chiral multiplets in
$\Omega$-deformed topologically twisted $\CN = (2,2)$ supersymmetric
field theories.

\section{\texorpdfstring{$\Omega$-deformations of $\CN = (2,2)$ supersymmetric field theories}{Ω-deformations of N = (2,2) supersymmetric field theories}}
\label{sec:2d}

As explained in the introduction, our approach to the chiral algebras
of four-dimensional $\CN = 2$ supersymmetric field theories is based
on $\Omega$-deformations in two dimensions.  In this section, we
formulate $\Omega$-deformations of topologically twisted $\CN = (2,2)$
supersymmetric field theories and SCFTs, and discuss some of their
properties that will be important in the application to
four-dimensional theories.

\subsection{\texorpdfstring{$\CN = (2,2)$ supersymmetry and superconformal symmetry}{N = (2,2) supersymmetry and superconformal symmetry}}

An $\CN = (2,2)$ supersymmetric field theory is a two-dimensional
quantum field theory that is invariant under a symmetry generated by
conserved charges forming the $\CN = (2,2)$ supersymmetry algebra.
Likewise, an $\CN = (2,2)$ SCFT is a theory whose symmetry algebra
contains the global $\CN = (2,2)$ superconformal algebra.

The global $\CN = (2,2)$ superconformal algebra is generated by the
generators $L_0$, $L_{\pm 1}$, $\Lb_0$, $\Lb_{\pm 1}$ of global
conformal transformations and $J_0$, $\Jb_0$ of $\U(1)$ R-symmetries,
Poincar\'e supercharges $G^\pm_{-\frac12}$, $\Gb^\pm_{-\frac12}$, and
conformal supercharges $G^\pm_{\frac12}$, $\Gb^\pm_{\frac12}$.  They satisfy
the commutation relations
\begin{equation}
  \label{eq:N=(2,2)SCA}
  \begin{alignedat}{2}
    [L_m,L_n] &= (m-n) L_{m+n} \,,
    &
    [\Lb_m,\Lb_n] &= (m-n) \Lb_{m+n} \,,
    \\
    [L_m, G^\pm_r] &= \Bigl(\frac{m}{2} - r\Bigr) G^\pm_{m+r} \,,
    &
    [\Lb_m, \Gb^\pm_r] &= \Bigl(\frac{m}{2} - r\Bigr) \Gb^\pm_{m+r} \,,
    \\
    [J_0, G^\pm_r] &= \pm G^\pm_r \,,
    &
    [\Jb_0, \Gb^\pm_r] &= \pm \Gb^\pm_r \,,
    \\
    \{G^+_r, G^-_s\} &= L_{r+s} + \frac{r-s}{2} J_{r+s} \,,
    &\qquad
    \{\Gb^+_r, \Gb^-_s\} &= \Lb_{r+s} + \frac{r-s}{2} \Jb_{r+s} \,,
  \end{alignedat}
\end{equation}
where $m$, $n = 0$, $\pm 1$ and $r$, $s = \pm \frac12$.  The other
combinations of the generators commute.  If we let $m$, $n$ run over
all integers and $r$, $s$ over all half-integers, then we obtain the
local $\CN = (2,2)$ superconformal algebra (with vanishing central
charges).

The supercharges are also denoted as
\begin{alignat}{4}
  Q_+ &= \Gb^-_{-\frac12} \,,
  &\qquad
  Q_- &= G^-_{-\frac12} \,,
  &\qquad
  \Qb_+ &= \Gb^+_{-\frac12} \,,  
  &\qquad
  \Qb_- &= G^+_{-\frac12} \,,  
  \\
  S_+ &= \Gb^-_{\frac12} \,,
  &\qquad
  S_- &= G^-_{\frac12} \,,
  &\qquad
  \Sb_+ &= \Gb^+_{\frac12} \,,  
  &\qquad
  \Sb_- &= G^+_{\frac12} \,.
\end{alignat}
In this notation, the subscripts $+$ and $-$ indicate that the
supercharges are from the left- and right-moving (or holomorphic and
antiholomorphic) sectors, respectively.  The supercharges are
characterized by their spin $M$, vector R-charge $F_V$ and axial
R-charge $F_A$, defined by
\begin{equation}
  M = 2(L_0 - \Lb_0) \,,
  \qquad
  F_V = J_0 + \Jb_0 \,,
  \qquad
  F_A = -J_0 + \Jb_0 \,.
\end{equation}
See Table~\ref{tab:q-numbers} for a summary of the spin and R-charges
of the supercharges.

\begin{table}
  \centering
  \begin{tabular}{c|cccc}
          & $Q_+$ & $Q_-$ & $\Qb_+$ & $\Qb_-$ \\ \hline
    $M$   & $-1$  & $+1$  & $-1$    & $+1$ \\
    $F_V$ & $-1$  & $-1$  & $+1$    & $+1$ \\
    $F_A$ & $-1$  & $+1$  & $+1$    & $-1$ \\
    $M_A$ & $-2$  & $0$   & $0$     & $+2$ \\
    $M_B$ & $-2$  & $+2$  & $0$     & $0$
  \end{tabular}
  \caption{The spin, R-charges and twisted spins of the supercharges
    in the $\CN = (2,2)$ supersymmetry algebra.}
  \label{tab:q-numbers}
\end{table}

The $\CN = (2,2)$ supersymmetry algebra is the subalgebra of the
$\CN = (2,2)$ superconformal algebra generated by $L_{-1}$,
$\Lb_{-1}$, $M$, $Q_\pm$, $\Qb_\pm$, $F_V$ and $F_A$.  The fact that
the $\U(1)_V \times \U(1)_A$ R-symmetry, generated by $F_V$ and $F_A$,
is unbroken implies that central charges do not appear in the algebra,
although it is straightforward to incorporate them.

The commutation relations~\eqref{eq:N=(2,2)SCA} are consistent with
the hermiticity condition
\begin{equation}
  \begin{alignedat}{2}
    (L_n)^\dagger &= L_{-n} \,,
    &
    (\Lb_n)^\dagger &= \Lb_{-n} \,,
    \\
    (G^\pm_r)^\dagger &= G^\mp_{-r} \,,
    &\qquad
    (\Gb^\pm_r)^\dagger &= \Gb^\mp_{-r} \,,
    \\
    (J_0)^\dagger &= J_0 \,,
    &
    (\Jb_0)^\dagger &= \Jb_0 \,.
  \end{alignedat}
\end{equation}
This is the action of hermitian conjugation in radial quantization.
The map
\begin{equation}
  \label{eq:A<->B-SC}
  G^\pm_r \mapsto G^\mp_r \,,
  \qquad
  J_0 \mapsto -J_0
\end{equation}
is an involution on the algebra.  On the $\CN = (2,2)$ supersymmetry
algebra it acts by the exchange
\begin{equation}
 \label{eq:A<->B-SUSY}
  Q_- \leftrightarrow \Qb_- \,,
  \qquad
  F_V \leftrightarrow F_A \,.
\end{equation}

\subsection{Topological twists}

A twist of a quantum field theory~\cite{Witten:1988ze, Witten:1988xj}
is an operation which changes the action of the rotation symmetry in
such a way that its generator $M$ is shifted to $M'$, differing by a
generator $F$ of a global symmetry:
\begin{equation}
  M \to M' = M + F \,.
\end{equation}
One way to think about this operation is that it promotes the global
symmetry to a local symmetry and identifies the associated gauge field
with a multiple of the spin connection.  Therefore, a twisted theory
is equivalent to the original theory if the spacetime is flat -- the
twist merely relabels some symmetry generators and changes what we
should call the energy--momentum tensor -- but the difference becomes
important otherwise.

For supersymmetric field theories, twists are particularly interesting
when the twisted theories have scalar supercharges.  The corresponding
fermionic symmetries are parametrized by constants rather than
covariantly constant spinors.  As such, typically these symmetries can
be realized on arbitrary manifolds.

Furthermore, in many cases one finds that a twisted theory has a
scalar supercharge $Q$ such that $Q^2 = 0$.  Then, one can consider
$Q$-cohomology in the space of operators and the space of states.  By
a standard argument, $Q$-invariant quantities in the twisted theory,
such as correlation functions of $Q$-closed operators, depend on the
objects involved only through their $Q$-cohomology classes.  Often,
the energy--momentum tensor is $Q$-exact and the diffeomorphisms act
trivially in the $Q$-cohomology.  In that situation, the $Q$-invariant
sector of the twisted theory defines a topological quantum field
theory.

In the case of $\CN = (2,2)$ supersymmetric field theories, there are
essentially two such topological twists.  In two dimensions, the
rotation group $\U(1)_M$ has a single generator $M$.  In the
\emph{A-twist}~\cite{Witten:1988xj}, the twisted rotation generator
$M'$ is given by
\begin{equation}
  \label{eq:M_A}
  M_A = M + F_V \,,
\end{equation}
and one considers the $Q$-cohomology with $Q$ being the supercharge
\begin{equation}
  Q_A = \Qb_+ + Q_- \,.
\end{equation}
(By these equations we mean that the generators on the left-hand sides
are identified with those on the right-hand sides when the spacetime
is flat.)  In the \emph{B-twist}~\cite{Vafa:1990mu, Witten:1991zz},
$M'$ is given by
\begin{equation}
  \label{eq:M_B}
  M_B = M + F_A \,,
\end{equation}
and $Q$ is taken to be
\begin{equation}
  Q_B = \Qb_+ + \Qb_- \,.
\end{equation}
For either twist, the generators $L_{-1}$, $\Lb_{-1}$ of translations
are $Q$-exact and annihilate $Q$-cohomology classes.  For theories
with local $\CN = (2,2)$ superconformal symmetry, the full topological
invariance, for an appropriate definition of the energy--momentum
tensor, follows from the superconformal algebra.

The $Q_A$-cohomology is graded by $F_A$, whereas the $Q_B$-cohomology
is graded by $F_V$.  The two twists are exchanged under the involution
\eqref{eq:A<->B-SUSY}:
\begin{equation}
  \label{eq:A<->B}
  M_A \leftrightarrow M_B \,,
  \qquad
  Q_A \leftrightarrow Q_B \,,
  \qquad
  F_V \leftrightarrow F_A \,.
\end{equation}

\subsection{\texorpdfstring{$\Omega$-deformations}{Ω-deformations}}

Consider a twisted supersymmetric field theory with a scalar
square-zero supercharge $Q$, and place it on a spacetime $\Sigma$ that
has a continuous isometry generated by a Killing vector field $V$.  An
\emph{$\Omega$-deformation}~\cite{Nekrasov:2002qd, Nekrasov:2003rj} of
the twisted theory on $\Sigma$ with respect to $V$ is a deformation
whereby $Q$ is replaced with a scalar supercharge $Q_V$ such that
\begin{equation}
  Q_V^2 = L_V \,.
\end{equation}
Here $L_V$ is the conserved charge corresponding to $V$; on fields,
$L_V$ acts as the Lie derivative with respect to $V$.  Slightly more
generally, we allow $V$ to be a complex linear combination of Killing
vector fields and define the $\Omega$-deformation with respect to it
by the same relation.

Since $Q_V$ no longer squares to zero, in order to define its
cohomology we must restrict $Q_V$ to $V$-invariant operators and
states.  Generically, this means that local operators must be located
at zeros of $V$.  This constraint is not a disadvantage compared to
the undeformed $Q$-cohomology if the generators of translations are
$Q$-exact, as the $Q$-cohomology class of a local operator does not
depend on the location then.  Rather, the lack of freedom can lead to
a richer algebraic structure, as we will see later.

For a topological twist of an $\CN = (2,2)$ supersymmetric field
theory, an $\Omega$-deformation may be constructed as
follows~\cite{Yagi:2014toa, Luo:2014sva}.

If $\Sigma$ is a flat surface, the twisted theory has a one-form
supercharge $\Qone = \Qone_\mu \rmd x^\mu$ such that
\begin{equation}
  \{Q,\Qone_\mu\} = \iu P_\mu \,,
  \qquad
  \{\Qone_\mu, \Qone_\nu\} = 0 \,,
\end{equation}
where $(x^1, x^2)$ are flat coordinates and $P_\mu$ are generators of
translations.  For $\Sigma = \C$, this supercharge equals
\begin{equation}
  \Qone_A = -\Qb_- \rmd z - Q_+ \rmd \zb
\end{equation}
for the A-twist and
\begin{equation}
  \Qone_B = -Q_- \rmd z - Q_+ \rmd \zb
\end{equation}
for the B-twist.
%
% Here we are using the representation of the Virasoro algebra (with
% $c = 0$) given by $L_n = -z^{n+1} \del_z$ and
% $\Lb_n = -z^{n+1} \del_\zb$.  We have $\iu P_\mu = \del_\mu$.
%
If $V = V^\mu \del_\mu$ has constant components
$V^\mu$, then $\iota_V \Qone = V^\mu \Qone_\mu$ is a conserved charge
and
\begin{equation}
  Q_V = Q + \iota_V \Qone
\end{equation}
satisfies $Q_V^2 = \iota_V P = L_V$.

After understanding how $Q_V$ transforms the fields on a flat
spacetime, we rewrite the transformation law so that it makes sense
even when $V$ is a vector field on a surface that is not necessarily
flat, and moreover it squares to $L_V$.  (Note that $\Qone_\mu$
generally do not exist, and even if they do, $L_V \neq \iota_V P_\mu$
unless $V^\mu$ are constants.)  Concretely, for vector and chiral
multiplets, we are then led to the formulas listed in
appendix~\ref{sec:QV-vector-chiral}.

If the transformation thus obtained is already a symmetry of the
twisted theory, the associated conserved charge serves as $Q_V$.  In
that event, the $\Omega$-deformation changes the cohomology but not
the underlying twisted theory.

In general, this is not what happens unless $V$ is covariantly
constant.  Yet, when the theory has a standard Lagrangian description
and $V$ is a Killing vector field, we can construct a new action
functional that is invariant under the $\Omega$-deformed
transformation.  Roughly speaking, in the new action the $Q$-exact
terms from the original action are replaced by the corresponding
$Q_V$-exact terms.  The $Q_V$-invariance relies on the fact that the
action of the twisted theory is constructed with the exterior
derivative and the Hodge star operator, acting on fields which are
differential forms.  Both of these operators commute with the Lie
derivative $L_V$ if $V$ generates an isometry, so $Q_V$ annihilates
the $Q_V$-exact terms thanks to the formula
$L_V = \rmd \iota_V + \iota_V \rmd$ and integration over $\Sigma$.

\subsection{\texorpdfstring{$\Omega$-deformation of $\CN = (2,2)$ SCFTs on $\R^2$}{Ω-deformation of N = (2,2) SCFTs on R2}}
\label{sec:Omega-SCFT}

Now suppose that the theory under consideration is an $\CN = (2,2)$
SCFT and defined on $\R^2$.  In this case, there is a canonical way to
make sense of $\iota_V \Qone$ itself as a conserved charge.
Therefore, the $\Omega$-deformation does not really deform twisted
$\CN = (2,2)$ SCFTs on $\R^2$; it only changes which cohomology we
should take.

This claim is clearly true if $V$ is a generator of translations, so
let us consider the case when it generates rotations.  Using the
angular coordinate $\theta$ on $\R^2$ we can write
\begin{equation}
  V = 2\iu\hbar\del_\theta \,,
\end{equation}
where we think of $\hbar$ as a complex parameter.  We denote the
corresponding $\Omega$-deformed supercharge $Q_V$ by $Q^\hbar$.  It
squares to the twisted rotation generator:
\begin{equation}
  \label{eq:Qhbar2}
  (Q^\hbar)^2 = \hbar M' \,.
\end{equation}

The definition of $\iota_V \Qone$ we are going to give works for all
$\CN = (2,2)$ SCFTs.  To motivate the definition, however, we first
look at theories with local $\CN = (2,2)$ superconformal symmetry,
such as free vector and free chiral multiplets.

For those theories, we can use a local conformal transformation to map
$\R^2$ to a cylinder $\R \times S^1$ plus a point at infinity at one
end.  Because $\R^2$ and $\R \times S^1$ are flat, on these geometries
the twisted and untwisted theories are naturally identified.

On the cylinder $V$ is just a generator of translations, so
$\iota_V\Qone$ can, and should, be defined as
$V^\mu \Qone_\mu = 2\iu\hbar \Qone_\theta$.  By the inverse conformal
transformation, $\Qone_\theta$ is mapped to a linear combination of
$G^\pm_r$ and $\Gb^\pm_r$, with $r \in \Z + \frac12$, in the theory on
$\R^2$.  Noting that neither $\Qone_\theta$ nor the conformal map
depends on $\hbar$, we can determine this linear combination by
comparing both sides of the relation~\eqref{eq:Qhbar2} order by order
in $\hbar$.  We find
\begin{equation}
  \label{eq:Qhbar}
  Q^\hbar = Q + 2\hbar S \,,
\end{equation}
with $S$ given by
\begin{align}
  \label{eq:S}
  S_A &= \Sb_- - S_+ \,,
  \\
  S_B &= S_- - S_+
  \label{eq:Q^hbar_B}
\end{align}
for the A-twist and the B-twist, respectively.

A possibly more direct derivation of the above formulas goes as
follows.  Recall that the fermionic generators of local $\CN = (2,2)$
superconformal symmetry are the modes of holomorphic supercurrents
$G^\pm$ and antiholomorphic supercurrents $\Gb^\pm$:
\begin{equation}
  G^\pm(z)
  = \sum_{r \in \Z + \frac12} \frac{G^\pm_r}{z^{r + \frac32}} \,,
  \qquad
  \Gb^\pm(\zb)
  = \sum_{r \in \Z + \frac12} \frac{\Gb^\pm_r}{\zb^{r + \frac32}} \,.
\end{equation}
In terms of these supercurrents, the one-form supercharge, say in the
B-twist, is given by
\begin{equation}
  \Qone_B
  =
  -\biggl(\oint \frac{\rmd z}{2\pi\iu} G^-\biggr) \rmd z
  + \biggl(\oint \frac{\rmd\zb}{2\pi\iu} \Gb^-\biggr) \rmd\zb
  \,,
\end{equation}
where the integration contours are circles of positive orientation
centered at the origin.  Then, for a vector field $V$ of the form
$V(z,\zb) = V^z(z) \del_z + V^\zb(\zb) \del_\zb$, it is natural to
define
\begin{equation}
  \iota_V \Qone_B
  =
  -\oint \frac{\rmd z}{2\pi\iu} V^z G^-
  + \oint \frac{\rmd\zb}{2\pi\iu} V^\zb \Gb^-
  \,.
\end{equation}
For $V = 2\iu\hbar(\iu z\del_z - \iu\zb\del_\zb)$, this reproduces the
formula $\iota_V \Qone_B = 2\hbar (S_- - S_+)$.

We see that the resulting formulas for $\iota_V\Qone$ for theories on
$\R^2$ only involve conformal supercharges in the global $\CN = (2,2)$
superconformal algebra.  Hence, these formulas also make sense for
theories with global but not local $\CN = (2,2)$ superconformal
symmetry, and can be employed as the definition of $\iota_V\Qone$ for
all $\CN = (2,2)$ SCFTs on $\R^2$.

An important observation about the $\Omega$-deformation of
$\CN = (2,2)$ SCFTs on $\R^2$, which follows from the
formula~\eqref{eq:Qhbar} for $Q^\hbar$, is that the
$Q^\hbar$-cohomologies for different values of $\hbar$ are all
isomorphic as long as $\hbar \neq 0$.  This is because $\hbar$ is
rescaled by a $\C^\times$-action, generated by $F_A$ in the A-twist
and by $F_V$ in the B-twist.  The $Q^\hbar$-cohomology is still graded
by $F_A$ or $F_V$ if we assign degree $2$ to $\hbar$.

\subsection{\texorpdfstring{$Q^\hbar$-cohomology of local operators for unitary
  $\CN = (2,2)$ SCFTs}{Qħ-cohomology of local operators for unitary
  N = (2,2) SCFTs}}
\label{sec:uSCFT}

The main object of interest in this paper is the $Q^\hbar$-cohomology
in the space of local operators that are located at the origin of
$\R^2$ and annihilated by $M'$.  We will refer to it simply as the
$Q^\hbar$-cohomology of local operators.

For unitary $\CN = (2,2)$ SCFTs, this $Q^\hbar$-cohomology has an
alternative description.  Let us write $Q^\hbar$ as
\begin{equation}
  Q^\hbar = \QQ^\hbar + \QQt^\hbar \,,
\end{equation}
where $\QQ^\hbar$ and $\QQt^\hbar$ satisfy
$(\QQ^\hbar)^2 = (\QQt^\hbar)^2 = 0$.  Specifically,
\begin{equation}
  \QQ_A^\hbar = \Qb_+ + 2\hbar \Sb_- \,,
  \qquad
  \QQt_A^\hbar = Q_- - 2\hbar S_+
\end{equation}
for the A-twist and
\begin{equation}
  \label{eq:QQB}
  \QQ_B^\hbar = \Qb_+ + 2\hbar S_- \,,
  \qquad
  \QQt_B^\hbar = \Qb_- - 2\hbar S_+
\end{equation}
for the B-twist.  Then, the $Q^\hbar$-cohomology of local operators is
isomorphic to the cohomology of local operators at the origin with
respect to either $\QQ^\hbar$ or $\QQt^\hbar$.

Here we prove the isomorphism for $|\hbar| = 1/4$.  Since the
$Q^\hbar$-cohomology is independent of $\hbar$, we have no loss of
generality.

First of all, note that we can restrict the action of the relevant
supercharges to the space of local operators that are located at the
origin and have $M' = 0$.  For $Q^\hbar$, this restriction is part of
the definition of the $Q^\hbar$-cohomology of local operators.  For
$\QQ^\hbar$ and $\QQt^\hbar$, we lose nothing since $M'$ is
$\QQ^\hbar$- and $\QQt^\hbar$-exact.

By conformal invariance, the space of local operators at the origin is
isomorphic as a vector space to the space of states on a circle.  A
standard argument about the relation between cohomology classes and
harmonic states then tells that the cohomology groups with respect to
$Q^\hbar$, $\QQ^\hbar$ and $\QQt^\hbar$ are isomorphic to the spaces
of local operators at the origin with
$\{Q^\hbar, (Q^\hbar)^\dagger\} = 0$,
$\{\QQ^\hbar, (\QQ^\hbar)^\dagger\} = 0$ and
$\{\QQt^\hbar, (\QQt^\hbar)^\dagger\} = 0$, respectively.  The
isomorphisms naturally extend to isomorphisms of rings.

The isomorphisms we wish to establish follow from the equalities
\begin{equation}
  \{Q^\hbar, (Q^\hbar)^\dagger\}
  = 2\{\QQ^\hbar, (\QQ^\hbar)^\dagger\}
  = 2\{\QQt^\hbar, (\QQt^\hbar)^\dagger\} \,.
\end{equation}

\subsection{\texorpdfstring{Localization of $\Omega$-deformed
    B-twisted gauge theories}{Localization of Ω-deformed B-twisted
    gauge theories}}
\label{sec:localization}

A salient feature of topologically twisted theories is localization of
the path integral to a simpler, often finite-dimensional, integration
over the space of supersymmetric configurations.  In the well-known
examples of the topological sigma models, the path integral localizes
to the space of holomorphic maps in the A-twisted
case~\cite{Witten:1988xj} and the space of constant maps in the
B-twisted case~\cite{Vafa:1990mu, Witten:1991zz}.

Similar localization takes place for $\Omega$-deformed twisted
theories.  Here we describe the localization formula for
$\Omega$-deformed B-twisted gauge theories~\cite{Yagi:2014toa,
  Luo:2014sva, Nekrasov:2018pqq, Costello:2018txb}.
See~\cite{Costello:2018txb} for a derivation and detailed discussions.

Consider an $\CN = (2,2)$ supersymmetric gauge theory, constructed
from a vector multiplet for gauge group $\SG$ and a chiral multiplet
in a representation $\SR\colon \SG \to \GL(\SV)$ of $\SG$.  The
superpotential $W$ of the theory is a gauge-invariant holomorphic
function on the representation space $\SV$, in which the complex
scalar field $\varphi$ of the chiral multiplet takes values.  We
perform the B-twist, place the twisted theory on $\R^2$, and turn on
the $\Omega$-deformation.

The path integral depends on the boundary condition at infinity.  For
the purpose of localization, a good boundary condition is specified as
follows.  Let $\Mod$ be the quotient of $\SV$ by the action of the
complexified gauge group $\SG_\C$.  (More precisely, we take $\Mod$ to
be the set of semistable $\SG_\C$-orbits.)  Since $W$ is
$\SG$-invariant and holomorphic, it is invariant under the
$\SG_\C$-action and defines a holomorphic function on $\Mod$.  The
moduli space $\Mod$ is K\"ahler, and so is the critical locus of $W$
in $\Mod$.  Choose a Lagrangian submanifold $\Lag_\infty$ of the
critical locus.  On the boundary, we require $\varphi$ to lie in
$\Lag_\infty$.

The localized path integral is then
\begin{equation}
  \int_\Lag \vol_\Lag
  \exp\biggl(-\frac{\iu\pi}{\hbar} W\biggr)
  \,,
\end{equation}
where $\vol_\Lag$ is the volume form on $\Lag$.  The integration
domain $\Lag$ is the union of all gradient flows generated by
$\Re(W/\hbar)$ in $\Mod$, terminating on $\Lag_\infty$.  This is a
Lagrangian submanifold of $\Mod$.

The above integral may be viewed as the path integral for a
zero-dimensional gauged sigma model with gauge group $\SG_\C$ and
target the preimage of $\Lag$ under the projection $\SV \to \Mod$.
The action function of this zero-dimensional theory is given by the
superpotential, and the $\Omega$-deformation parameter plays the role
of the Planck constant.

\section{\texorpdfstring{Chiral algebras from
    $\Omega$-deformation}{Chiral algebras from Ω-deformation}}
\label{sec:4d}

Now we apply what we have learned in the previous section to $\CN = 2$
supersymmetric field theories in four dimensions.  The goal of this
section is to understand how chiral algebras arise from these theories
via the combination of the topological--holomorphic twist and
$\Omega$-deformation, and how they are related to the chiral algebras
introduced in~\cite{Beem:2013sza}.

\subsection{\texorpdfstring{$\CN = 2$ supersymmetry and superconformal
    symmetry}{N = 2 supersymmetry and superconformal symmetry}}

To begin with, we review basic facts about $\CN = 2$ supersymmetry and
superconformal symmetry in four dimensions.  We refer the reader
to~\cite{Beem:2013sza} for more details.

The $\CN = 2$ superconformal algebra is generated by the generators
$\CP_{\alpha\dot\alpha}$ of translations, $\CM_\alpha{}^\beta$,
$\CM^{\dot\alpha}{}_{\dot\beta}$ of rotations, $\CH$ of dilatations,
$\CK^{\dot\alpha\alpha}$ of special conformal transformations,
$\CR^\CI{}_\CJ$ of the R-symmetry group
$\U(2)_\CR \iso \SU(2)_R \times \U(1)_r$, as well as eight Poincar\'e
supercharges $\CQ^\CI_\alpha$, $\CQt_{\CI\dot\alpha}$ and eight
conformal supercharges $\CS_\CI^\alpha$, $\CSt^{\CI\dot\alpha}$.  Here
$\CI$, $\CJ = 1$, $2$ are indices for $\U(2)_\CR$ doublets, and
$\alpha$, $\beta = \pm$ and $\dot\alpha$, $\dot\beta = \dot\pm$ are
indices for Weyl spinors.

The $\CN = 2$ supersymmetry algebra is the subalgebra of the $\CN = 2$
superconformal algebra generated by $\CP_{\alpha\dot\alpha}$,
$\CM_\alpha{}^\beta$, $\CM^{\dot\alpha}{}_{\dot\beta}$,
$\CR^\CI{}_\CJ$, $\CQ^\CI_\alpha$, $\CQt_{\CI\dot\alpha}$.  The
unbroken $\U(1)_r$ symmetry implies that the central charge vanishes.

We list some of the commutation relations in the $\CN = 2$
superconformal algebra which will be of particular importance in the
subsequent discussions.  The nonvanishing commutators between the
supercharges are
\begin{equation}
  \begin{aligned}
    \{\CQ^\CI_\alpha, \CQt_{\CJ\dot\alpha}\}
    &= \delta^\CI_\CJ \CP_{\alpha\dot\alpha} \,,
    \\
    \{\CSt^{\CI\dot\alpha}, \CS_\CJ^\alpha\}
    &= \delta^\CI_\CJ \CK^{\dot\alpha\alpha} \,,
    \\
    \{\CQ^\CI_\alpha, \CS_\CJ^\beta\}
    &= \frac12 \delta^\CI_\CJ \delta_\alpha^\beta \CH
    + \delta^\CI_\CJ \CM_\alpha{}^\beta
    - \delta_\alpha^\beta \CR^\CI{}_\CJ \,,
    \\
    \{\CSt^{\CI\dot\alpha}, \CQt_{\CJ\dot\beta}\}
    &= \frac12 \delta^\CI_\CJ \delta^{\dot\alpha}{}_{\dot\beta} \CH
    + \delta^\CI_\CJ \CM^{\dot\alpha}{}_{\dot\beta}
    + \delta^{\dot\alpha}_{\dot\beta} \CR^\CI{}_\CJ \,.
  \end{aligned}
\end{equation}
Rotations act on the Poincar\'e supercharges as
\begin{equation}
  \begin{aligned}[]
    [\CM_\alpha{}^\beta, \CQ^\CI_\gamma]
    &= \delta_\gamma^\beta \CQ^\CI_\alpha
    - \frac12 \delta_\alpha^\beta \CQ^\CI_\gamma \,,
    \\
    [\CM^{\dot\alpha}{}_{\dot\beta}, \CQt_{\CI\dot\gamma}]
    &= \delta^{\dot\alpha}_{\dot\gamma} \CQt_{\CI\dot\beta}
    - \frac12 \delta^{\dot\alpha}_{\dot\beta} \CQt_{\CI\dot\gamma} \,,
    % \\
    % [\CM_\alpha{}^\beta, \CS_\CI^\gamma]
    % &= -\delta_\alpha^\gamma \CS_\CI^\beta
    % + \frac12 \delta_\alpha^\beta \CS_\CI^\gamma \,,
    % \\
    % [\CM^{\dot\alpha}{}_{\dot\beta}, \CSt^{\CI\dot\gamma}]
    % &= -\delta^{\dot\gamma}_{\dot\beta} \CSt^{\CI\dot\alpha}
    % + \frac12 \delta^{\dot\alpha}_{\dot\beta} \CSt^{\CI\dot\gamma} \,,
  \end{aligned}
\end{equation}
whereas R-symmetry transformations act on them as
\begin{equation}
  \begin{aligned}[]
    [\CR^\CI{}_\CJ, \CQ^\CK_\alpha]
    &= \delta_\CJ^\CK \CQ^\CI_\alpha
    - \frac14 \delta^\CI_\CJ \CQ^\CK_\alpha \,,
    \\
    [\CR^\CI{}_\CJ, \CQt_{\CK\dot\alpha}]
    &= -\delta^\CI_\CK \CQt_{\CJ\dot\alpha}
    + \frac14 \delta^\CI_\CJ \CQt_{\CK\dot\alpha} \,.
    % \\
    % [\CR^\CI{}_\CJ, \CS_\CK^\alpha]
    % &= -\delta^\CI_\CK \CS_\CJ^\alpha
    % + \frac14 \delta^\CI_\CJ \CS_\CK^\alpha \,,
    % \\
    % [\CR^\CI{}_\CJ, \CSt^{\CK\dot\alpha}]
    % &= \delta^\CK_\CJ \CSt^{\CI\dot\alpha}
    % - \frac14 \delta^\CI_\CJ \CSt^{\CK\dot\alpha}
    % \,.
  \end{aligned}
\end{equation}
Their action on the conformal supercharges may be found from the 
hermiticity condition in radial quantization:
\begin{equation}
  \begin{gathered}
    \CH^\dagger = \CH \,,
    \qquad
    (\CP_{\alpha\dot\alpha})^\dagger = \CK^{\dot\alpha\alpha} \,,
    \qquad
    (\CM_\alpha{}^\beta)^\dagger = \CM_\beta{}^\alpha \,,
    \qquad
    (\CM^{\dot\alpha}{}_{\dot\beta})^\dagger
    = \CM^{\dot\beta}{}_{\dot\alpha}\,,
    \\
    (\CR^\CI{}_\CJ)^\dagger = \CR^\CJ{}_\CI \,,
    \qquad
    (\CQ^\CI_\alpha)^\dagger = \CS_\CI^\alpha \,,
    \qquad
    (\CQt_{\CI\dot\alpha})^\dagger = \CSt^{\CI\dot\alpha} \,.
  \end{gathered}
\end{equation}

\subsection{Kapustin's topological--holomorphic twist}

In~\cite{Kapustin:2006hi}, Kapustin introduced a twist of an $\CN = 2$
supersymmetric field theory which is applicable when the theory is
placed on a product $\Sigma \times C$ of two surfaces.  Upon this
twist the theory becomes topological on $\Sigma$ and holomorphic on
$C$.  For this reason, Kapustin's twist is called a
topological--holomorphic twist.

Let us choose a local frame on the spinor bundle on $\Sigma \times C$
in such a way that
\begin{equation}
  M_\Sigma = 2(\CM_+{}^+ - \CM^{\dot+}{}_{\dot+})
\end{equation}
is a generator of rotations on $\Sigma$, while
\begin{equation}
  M_C = 2(\CM_+{}^+ + \CM^{\dot+}{}_{\dot+}) \,
\end{equation}
generates rotations on $C$.  The topological--holomorphic twist
replaces these generators with
\begin{align}
  M_\Sigma' &= M_\Sigma + 2r \,,
  \\
  M_C' &= M_C + 2R \,,
\end{align}
where we have introduced $r$ and $R$ by
\begin{equation}
  \CR^1{}_1 = \frac{r}{2} + R \,,
  \qquad
  \CR^2{}_2 = \frac{r}{2} - R \,.
\end{equation}
The R-charges $r$ and $R$ are generators of $\U(1)_r$ and the diagonal
subgroup $\U(1)_R$ of $\SU(2)_R$, respectively.
Table~\ref{tab:Kapustin-twist} summarizes how the supercharges
transform under the action of these $\U(1)$ charges.

\begin{table}
  \centering
  \begin{tabular}{c|cccccccc}
    \vphantom{$\bar{\CQt}$} &
    $\CQ^1_+$ & $\CQ^1_-$ & $\CQt_{1\dot+}$ & $\CQt_{1\dot-}$
    &
    $\CQ^2_+$ & $\CQ^2_-$ & $\CQt_{2\dot+}$ & $\CQt_{2\dot-}$
    \\
    \hline
    $M_\Sigma$
    &
    $+1$ & $-1$ & $-1$ & $+1$
    &
    $+1$ & $-1$ & $-1$ & $+1$
    \\
    $M_C$
    &
    $+1$ & $-1$ & $+1$ & $-1$
    &
    $+1$ & $-1$ & $+1$ & $-1$
    \\
    $2r$
    &
    $+1$ & $+1$ & $-1$ & $-1$
    &
    $+1$ & $+1$ & $-1$ & $-1$
    \\
    $2R$
    &
    $+1$ & $+1$ & $-1$ & $-1$
    &
    $-1$ & $-1$ & $+1$ & $+1$
    \\
    $M_\Sigma'$
    &
    $+2$ & $0$ & $-2$ & $0$
    &
    $+2$ & $0$ & $-2$ & $0$
    \\
    $M_C'$
    &
    $+2$ & $0$ & $0$ & $-2$
    &
    $0$ & $-2$ & $+2$ & $0$
  \end{tabular}
  \caption{The spins, R-charges and twisted spins of the Poincar\'e
    supercharges.  Those of the conformal supercharges are opposite of
    their hermitian conjugates'.}
  \label{tab:Kapustin-twist}
\end{table}

The twisted theory has two scalar supercharges $\CQ^1_-$ and
$\CQt_{2\dot-}$.  Both of these square to zero and they commute with
each other, so we can consider the cohomology with respect to the
linear combination
\begin{equation}
  Q = \CQ^1_- + t \CQt_{2\dot-}
\end{equation}
for any $t \in \C \cup \{\infty\}$.  However, the cohomologies for all
values of $t$ other than $0$ or $\infty$ are isomorphic via the
$\C^\times$-action generated by $r$.  It is convenient to set
\begin{equation}
  t = 1 \,.
\end{equation}

The momenta $\CP_{+\dot-}$, $\CP_{-\dot+}$ and $\CP_{-\dot-}$ are
$Q$-exact.  From their commutators with $M_\Sigma'$ and $M_C'$, one
finds that the first two are generators of translations on $\Sigma$,
whereas the last one is a generator of antiholomorphic translations on
$C$.  Therefore, $Q$-cohomology classes of local operators are
independent of their positions on $\Sigma$ and depend holomorphically
on $C$.  For a gauge theory with a standard Lagrangian formulation,
one can moreover show that the action is independent of the metric on
$\Sigma$, up to $Q$-exact terms.  In this sense, the twisted theory is
topological on $\Sigma$ and holomorphic on $C$.  We will call it a
topological--holomorphic theory on $\Sigma \times C$.

\subsection{\texorpdfstring{$\Omega$-deformed topological--holomorphic
    theories and chiral algebras}{Ω-deformed
    topological-holomorphic theories and chiral algebras}}
\label{sec:CA-Omega}

As it is, the algebra of local operators of the
topological--holomorphic theory is not very interesting because the
product of two local operators does not contain any singularities.
This is a consequence of Hartogs's extension theorem, which states
that for $n \geq 2$, a holomorphic function on $U \setminus V$ can be
extended to a holomorphic function on $U$, where $U$ is an open subset
of $\C^n$ and $V$ is a compact subset of $U$ such that $U \setminus V$
is connected.

The problem is that the topological--holomorphic theory has the
topological surface $\Sigma$ on which local operators can move freely.
In order to obtain an operator product structure that allows
singularities, we must make a modification to the
topological--holomorphic theory that eliminates this freedom and turns
local operators into holomorphic functions of a single variable.

For $\Sigma = \R^2$, this is precisely what an $\Omega$-deformation
does: it requires local operators to stay at the origin.  Thus, via
the $\Omega$-deformation, the topological--holomorphic theory on
$\R^2 \times C$ produces a \emph{chiral algebra} on $C$, that is, the
algebra of local operators in a holomorphic field theory on $C$.

More precisely, this $\Omega$-deformation of the
topological--holomorphic theory for $\Sigma = \R^2$ is parametrized by
$\hbar \in \C$, and deforms $Q$ to $Q^\hbar$ satisfying
\begin{equation}
  \label{eq:QKhbar2}
  (Q^\hbar)^2 = \hbar M_\Sigma' \,.
\end{equation}
We assume that the generator of holomorphic translations on $C$ remains
$Q^\hbar$-closed and that of antiholomorphic translations
$Q^\hbar$-exact.  (The generators, however, may be shifted from
$\CP_{+\dot+}$ and $\CP_{-\dot-}$.)  Under this assumption, the
$Q^\hbar$-cohomology, taken in the space of rotation invariant local
operators placed at the origin of $\R^2$, defines the chiral algebra
of the theory.

To understand this construction better, let us describe the
topological--holomorphic twist in two steps.  In the first step we
perform a twist along $C$, replacing $M_C$ with $M_C'$.  After that,
we twist along $\Sigma$.

The first step turns the supercharges $\CQ^1_-$ $\CQt_{1\dot+}$,
$\CQ^2_+$, $\CQt_{2\dot-}$ into scalars on $C$, thereby enabling them
to be unbroken even when $C$ is curved.  Two of them have
$M_\Sigma = +1$ and the other two have $M_\Sigma = -1$.  Together with
$\CP_{+\dot-}$, $\CP_{-\dot+}$, $M_\Sigma$, $r$ and $R$, these
supercharges generate two-dimensional supersymmetry, namely the
$\CN = (2,2)$ supersymmetry algebra on $\Sigma$.

The relation between the four-dimensional generators and the
two-dimensional ones can be determined as follows.  Suppose that the
theory contains a vector multiplet.  If we twist and reduce the theory
on $C$, we obtain an $\CN = (2,2)$ supersymmetric theory on $\Sigma$.
In this process, the two real scalars in the vector multiplet become
the two real scalars in an $\CN = (2,2)$ vector multiplet.%
\footnote{Here we are choosing to describe the two-dimensional theory
  using a vector multiplet as opposed to a twisted vector multiplet.
  This is different from the choice made in \cite{Cordova:2017mhb,
    Pan:2017zie}.  As a result, our identification of the two- and
  four-dimensional supercharges differs from theirs by the
  involution~\eqref{eq:A<->B-SUSY}.}
Since complex linear combinations of the former scalars have
$(R,r) = (0,\pm 1)$, while complex linear combinations of the latter
have $(F_V, F_A) = (0,\pm 2)$, we learn that $F_A = \pm 2r$ up to
generators that act trivially on these scalars.  The choice of the
sign is a matter of convention, so let us pick $+$.  Requiring that
two supercharges have $F_A = +1$ and the other two have $F_A = -1$, we
find $F_A = 2r + \alpha M_C'$ for some constant $\alpha$.  Looking at
the values of $M_\Sigma$ and $F_A$ for the supercharges, we deduce the
identification
\begin{equation}
  Q_+ = \CQt_{1\dot+} \,,
  \qquad
  Q_- = \CQ^2_+ \,.
  \qquad
  \Qb_+ = \CQ^1_- \,,
  \qquad
  \Qb_- = \CQt_{2\dot-} \,,
\end{equation}
from which it also follows that $F_V = 2R + \beta M_C'$ for some
$\beta$.  Finally, from the commutation relations of the supercharges
we get
\begin{equation}
  L_{-1} = \CP_{+\dot-} \,,
  \qquad
  \Lb_{-1} = \CP_{-\dot+} \,.
\end{equation}

As far as the $\CN = (2,2)$ supersymmetry algebra is concerned, we can
always shift $\U(1)$ R-charges by generators of global $\U(1)$
symmetries commuting with the supercharges.  Thus, we simply postulate
$F_A = 2r$; as we will see in section~\ref{sec:CA-SCFT}, this relation
is consistent with what we find in the superconformal case.  Then, the
second step in the topological--holomorphic twist replaces $M_\Sigma$
with $M_\Sigma' = M_\Sigma + F_A$.  This is the B-twist of the
$\CN = (2,2)$ supersymmetry on $\Sigma$.  With our choice $t = 1$, the
supercharge used for the cohomology is
\begin{equation}
  \QK =  \CQ^1_- + \CQt_{2\dot-}
\end{equation}
and coincides with the B-twist supercharge $Q_B$.

The above consideration shows that the relevant $\Omega$-deformation
is a four-dimensional counterpart of the one for B-twisted theories in
two dimensions.  For theories constructed from vector multiplets and
hypermultiplets, one may obtain formulas for the $\Omega$-deformation
by lifting the two-dimensional formulas listed in
appendix~\ref{sec:QV-vector-chiral} to four dimensions.  A similar
lift was considered in~\cite{Costello:2018txb} in the context of a
topological--holomorphic twist of the six-dimensional maximally
supersymmetric Yang--Mills theory.

\subsection{\texorpdfstring{Chiral algebras of $\CN = 2$ SCFTs}{Chiral
    algebras of N = 2 SCFTs}}
\label{sec:CA-SCFT}

If the theory is superconformal and placed on
$\Sigma \times C = \R^2 \times \C$, we do not only get $\CN = (2,2)$
supersymmetry on $\Sigma$.  In this situation, we actually get global
$\CN = (2,2)$ superconformal symmetry.  Indeed, the conformal
supercharges $\CS^1_-$ $\CSt_{1\dot+}$, $\CS^2_+$, $\CSt_{2\dot-}$,
which are scalars with respect to $M_C'$, can be identified with the
two-dimensional conformal supercharges as
\begin{equation}
  S_+ = \CS_1^- \,,
  \qquad
  S_- = \CSt^{2\dot-} \,.
  \qquad
  \Sb_+ = \CSt^{1\dot+} \,,
  \qquad
  \Sb_- = \CS_2^+ \,.
\end{equation}
The identification of the remaining generators are as follows:
\begin{equation}  
  \begin{alignedat}{2}
    L_0 &= \frac12 \CH + \frac14 M_\Sigma \,,
    &\qquad
    \Lb_0 &= \frac12 \CH - \frac14 M_\Sigma \,,
    \\
    L_1 &= \CK^{\dot-+} \,,
    &
    \Lb_1 &= \CK^{\dot+-} \,,
    \\
    J_0 &= \frac12 M_C' - r + R \,,
    &
    \Jb_0 &= \frac12 M_C' + r + R \,.
  \end{alignedat}
\end{equation}
In particular, we have
\begin{equation}
  F_V = 2R + M_C' \,,
  \qquad
  F_A = 2r \,.
\end{equation}
The hermiticity conditions are consistent in two and four dimensions.

Since the theory has $\CN = (2,2)$ superconformal symmetry on $\R^2$,
the $\Omega$-deformation of the topological--holomorphic theory can be
achieved by the procedure described in section~\ref{sec:Omega-SCFT}.
As explained there, in the superconformal case the
$\Omega$-deformation just changes the supercharge with respect to
which the cohomology is defined.  According to the
formula~\eqref{eq:Q^hbar_B}, the $\Omega$-deformed supercharge is
\begin{equation}
  \QK^\hbar
  =
  \CQ^1_- + \CQt_{2\dot-} + 2\hbar (\CSt^{2\dot-} - \CS_1^-) \,,
\end{equation}
which indeed satisfies the relation~\eqref{eq:QKhbar2}.

For a unitary SCFT, the chiral algebra defined here is isomorphic to
the one introduced in \cite{Beem:2013sza}.  As we have seen in
section~\ref{sec:uSCFT}, in the unitary case the
$\QK^\hbar$-cohomology of local operators is isomorphic to the
cohomology of local operators at the origin with respect to either
$\QQK^\hbar$ or $\QQtK^\hbar$.  In the present case we have
\begin{equation}
  \QQK^\hbar = \CQ^1_- + 2\hbar \CSt^{2\dot-} \,,
  \qquad
  \QQtK^\hbar = \CQt_{2\dot-} - 2\hbar \CS_1^- \,.
\end{equation}
These are the supercharges used in the construction of the chiral
algebra in~\cite{Beem:2013sza}.

The generator $\CP_{+\dot+}$ of holomorphic translations on $C$ is
$\QK^\hbar$-closed, which is part of the assumption we have made.  The
generator of antiholomorphic translations is $\QK^\hbar$-exact,
provided that it is shifted from $\CP_{-\dot-}$ to
$\CP_{-\dot-} + 2\hbar \CR^2{}_1$:
\begin{equation}
  \{\QK^\hbar, \CQt^{1\dot-}\}
  = \{\QK^\hbar, \CQ^2_-\}
  = \CP_{-\dot-} + 2\hbar \CR^2{}_1 \,.
\end{equation}
Since $\CQt^{1\dot-}$ and $\CQ^2_-$ are the only fermionic generators
with $(M_\Sigma', M_C') = (0,-2)$, there is no other candidate for a
$\QK^\hbar$-exact generator of antiholomorphic translations.

In fact, $\CP_{-\dot-} + 2\hbar \CR^2{}_1$ is the only bosonic
generator of the $\CN = 2$ superconformal algebra such that it
commutes with $\QK^\hbar$, reduces to $\CP_{-\dot-}$ for $\hbar = 0$,
and has the correct values $(M_\Sigma', M_C') = (0,-2)$ and
$(F_V, F_A) = (-2,0)$ to represent $\del_\wb$ up to an overall factor.
At first order in~$\hbar$, the correction to $\CP_{-\dot-}$ is given
by a bosonic generator of the $\CN = 2$ supersymmetry algebra that has
$(M_\Sigma', M_C') = (0,-2)$ and $(F_V, F_A) = (-4,0)$.  (Recall that
$\hbar$ has $F_V = 2$ in the B-twist.) The only such generator is
$\CR^2{}_1$, and the commutativity with $\QK^\hbar$ fixes the
coefficient of the first order correction.  Higher order corrections
are absent as there are no generators with $F_V \leq -6$.  By the same
token, $\CP_{+\dot+}$ is the only $\QK^\hbar$-closed bosonic generator
that can represent $\del_w$.  This argument shows that if we use a set
of field variables on which the action of $\del_w$ and $\del_\wb$
commute with $\QK^\hbar$, then generically, $\CP_{+\dot+}$ and
$\CP_{-\dot-} + 2\hbar \CR^2{}_1$ generate translations along $C$.

\subsection{Vector multiplets and hypermultiplets}
\label{sec:VM-HM}

As an example, let us consider an $\CN = 2$ supersymmetric gauge
theory constructed from a vector multiplet for a gauge group $G$ and a
hypermultiplet in a representation $R_H\colon G \to \GL(V)$ of $G$.
For this theory, the chiral algebra arising from the
$\Omega$-deformation of the topological--holomorphic twist has been
found by Butson~\cite{Butson:2019} to be isomorphic to the one defined
in~\cite{Beem:2013sza}.

Here we derive the same result by localization.  The idea is to view
the topological--holomorphic theory on $\R^2 \times C$ as a B-twisted
gauge theory on $\R^2$, and apply the localization formula explained
in section~\ref{sec:localization}.

The gauge group $\SG$ of this two-dimensional theory is the space of
maps from $C$ to $G$; its elements are four-dimensional gauge
transformations that are constant on $\R^2$.  The $\CN = 2$ vector
multiplet splits into an $\CN = (2,2)$ vector multiplet and an
$\CN = (2,2)$ chiral multiplet in the adjoint representation of $\SG$.
The complex scalar in the adjoint chiral multiplet is $A_\wb$, which
is annihilated by $\QK$ and has $(F_V, F_A) = (-2,0)$.

The $\CN = 2$ hypermultiplet consists of a pair of $\CN = 1$ chiral
multiplets valued in $R_H$ and its dual $R_H^\vee$.  From the
two-dimensional point of view, it is a pair of $\CN = (2,2)$ chiral
multiplets valued in $\SR_H$ and $\SR_H^\vee$, where
$\SR_H\colon \SG \to \GL(\SV)$ is the representation of $\SG$ induced
from $R_H$ on the space $\SV$ of maps from $C$ to $V$.  The complex
scalars $q$ and $\qt$ of these multiplets have $(F_V, F_A) = (2,0)$.
Under the topological--holomorphic twist, they become spinors on $C$
with $M_C' = 1$, that is, sections of $K_C^{1/2}$.

If we wish to place the hypermultiplet on any choice of $C$, it is
necessary to further twist the theory along $C$ so that all fields
become differential forms rather than spinors.  For this twist we can
use the $\U(1)$ symmetry which commutes with the supercharges and
assigns $q$ and $\qt$ opposite charges.  We choose to make $q$ and
$\qt$ have $M_C' = 0$ and $2$, respectively.  To emphasize this change
we rename them as
\begin{equation}
  \label{eq:beta=qt}
  \beta = \qt \,,
  \qquad
  \gamma = q \,.
\end{equation}
Thus, $\beta$ is a $(1,0)$-form valued in $R_H^\vee$ and $\gamma$ is a
scalar valued in $R_H$ on $C$.

In order to apply the localization formula, we need to determine the
superpotential $W$ of the two-dimensional theory.  In the present
setup, $W$ is a gauge-invariant holomorphic functional of
$(A_\wb, q, \qt)$ with $(F_V, F_A) = (2,0)$.  Furthermore, it must be
the integral over $C$ of a local density which is of first order in
derivatives on $C$ so that the Lagrangian is local and of second
order.  The only such functional, up to an overall factor, is
\begin{equation}
  W = \int_C \beta \delb_A \gamma \,,
\end{equation}
where $\delb_A = \rmd\wb (\del_\wb + A_\wb)$ is the gauge covariant
Dolbeault operator on $C$.  The prefactor is irrelevant as it can be
absorbed by a rescaling of $\beta$ and $\gamma$.

Now we turn on the $\Omega$-deformation.  According to the
localization formula, the $\QK^\hbar$-invariant sector of the
$\Omega$-deformed B-twisted gauge theory on $\R^2$ is equivalent to a
zero-dimensional gauge theory whose gauge group is $\SG_\C$.  The
action of the theory (including the Planck constant) is $W/\hbar$.  In
the case at hand, this is a functional of fields that are maps from
$C$ to $V$, $V^\vee$ or the Lie algebra $\gf_\C$ of $G_\C$.  Hence,
the localized theory is really a quantum field theory on $C$.  The
algebra of local operators of this two-dimensional theory is the
chiral algebra in question.

The chiral algebra does not depend on the global structures of $C$, so
let us take $C = \C$.  Then, by a gauge transformation we can set%
\footnote{At first sight it may appear that holomorphic gauge
  transformations leave the gauge fixing condition intact.  This is
  not the case because $A$ is the gauge field for the compact gauge
  group $G$ and obeys the hermiticity condition
  $A_w = -(A_\wb)^\dagger$.  The residual gauge symmetry is therefore
  given by the constant gauge transformations.  This is a global
  symmetry, with respect to which we do not quotient out the field
  space.  Accordingly, we remove the constant gauge transformations
  from the BRST procedure, which means that the ghost $c$ does not
  contain the zero mode.  The ghosts without the zero mode of $c$ form
  the so-called small $bc$ system~\cite{Friedan:1985ge}.}
\begin{equation}
  A_\wb = 0 \,.
\end{equation}
The corresponding ghost action is
\begin{equation}
  \frac{1}{\hbar}
  \int_C \Tr\bigl(b \wedge \delb_A c + \auxB \wedge A\bigr) \,,
\end{equation}
where $c$ is a scalar and $b$, $\auxB$ are $(1,0)$-forms on $C$, all
valued in the adjoint representation of $G$.  Integrating out the
auxiliary field $\auxB$ produces a delta function in $A_\wb$ which
imposes the gauge fixing condition, while the integration over $b$ and
$c$ gives the associated Faddeev--Popov determinant.

Instead of integrating $\auxB$ out, let us first integrate over
$A_\wb$ to set
\begin{equation}
  \auxB = -\eta^{ab} \beta_i (T_b)^i{}_j \gamma^j T_a  + \{b,c\} \,.
\end{equation}
Here we have chosen a basis $\{T_a\}_{a=1}^{\dim\gf}$ for $\gf$ and
denoted by $\eta^{ab}$ the $(a,b)$ component of the inverse of the
Killing form; also, we have used $i$, $j$ for indices for the
representations $R_H$ and $R_H^\vee$.  The action of the localized
theory becomes
\begin{equation}
  \frac{1}{\hbar} \int_C (\beta \delb\gamma + b \delb c) \,.
\end{equation}
As explained in section \ref{sec:Omega-SCFT}, $\hbar$ can be rescaled
by the $\C^\times$-action generated by $F_V$, and we see this property
reflected in the fact that the action is quadratic.  We made use of
this rescaling when we identified the superpotential.

The chiral algebra described by the above action is the gauged
$\beta\gamma$ system, which has the operator product expansions (OPEs)
\begin{align}
  \beta_j(w) \gamma^i(w')
  &\sim -\frac{\hbar \delta^i_j \rmd w}{w - w'} \,,
  \\
  b^a(w) c^b(w')
  &\sim \frac{\hbar \eta^{ab} \, \rmd w}{w - w'}
\end{align}
and the BRST charge
\begin{equation}
  Q_\BRST
  =
  \frac{1}{2\pi\iu \hbar} \oint \bigl(-\beta c \gamma + \Tr(bcc)\bigr) \,.
\end{equation}
Essentially the same result was derived by localization computations
on $S^4$~\cite{Pan:2017zie} and $S^3 \times S^1$~\cite{Pan:2019bor,
  Dedushenko:2019yiw}.

Computing the double contractions in the OPE of two BRST currents, we
find an anomaly in the nilpotency of $Q_\BRST$:
\begin{equation}
  Q_\BRST^2
  =
  \bigl(\Tr_\adj(T_a T_b) - \Tr_{R_H}(T_a T_b)\bigr)
  \del_w c^a c^b \,,
\end{equation}
where $\Tr_{\mathrm{adj}}$ and $\Tr_{R_H}$ denote trace taken in the
adjoint representation and in $R_H$, respectively.  The factor in the
parentheses on the right-hand side vanishes precisely when $\U(1)_r$
is nonanomalous.  Indeed, under the action of
$e^{2\iu\alpha r} \in \U(1)_r$, the fermionic path integral measure
changes by the phase factor
\begin{equation}
  \label{eq:U(1)r-anomaly}
  \exp\biggl(-\frac{\iu\alpha}{2\pi^2}
  \int_{\R^2 \times C}  \bigl(\Tr_\adj(T_a T_b) - \Tr_{R_H}(T_a T_b)\bigr)
  F^a \wedge F^b\biggr) \,.
\end{equation}
The appearance of the $\U(1)_r$ anomaly is natural from the
perspective of the topological--holomorphic twist; as $\U(1)_r$ is
used in the twist, it had better be well defined.  The same condition
also dictates the vanishing of the one-loop beta function of the
theory.  From the point of view of the untwisted theory on $\R^4$,
this is necessary because the construction of~\cite{Beem:2013sza}
fails if the conformal symmetry is broken.

We remark that a care must be taken when we identify the fields in the
chiral algebra and those in the underlying $\CN = 2$ SCFT.  In fact,
the gauged $\beta\gamma$ system is written in terms of field variables
on which the antiholomorphic derivative $\del_\wb$ on $C$ is
represented by the $\QK^\hbar$-exact operator
$\CP_{-\dot-} + 2\hbar \CR^2{}_1$, not $\CP_{-\dot-}$.  The reason is
that in describing the $\CN = 2$ SCFT as a two-dimensional theory, we
have treated the coordinates on~$C$ as ``flavor indices'' so that the
theory is formulated as an $\Omega$-deformed $\CN = (2,2)$
supersymmetric gauge theory in the standard manner, with integration
over $C$ taking the place of summation over the flavor indices.  In
particular, we have demanded that $\del_\wb$ commutes with
$\QK^\hbar$, which requires the relation between $\del_\wb$ and
$\CP_{-\dot-}$ to be corrected, as explained at the end of
section~\ref{sec:CA-SCFT}.  As a result, the
identification~\eqref{eq:beta=qt} is modified in the $\Omega$-deformed
setting by conjugation by $\exp(2\hbar\wb \CR^2{}_1)$:
\begin{equation}
  \begin{aligned}
    \beta
    =
    \qt - 2\hbar \wb q^\dagger \,,
    \qquad
    \gamma
    = q + 2\hbar \wb \qt^\dagger \,.
  \end{aligned}
\end{equation}
These are called twisted-translated operators in~\cite{Beem:2013sza}.
The $\hbar$-corrections with explicit position dependence are crucial
for reproducing the correct OPEs within the $\CN = 2$ SCFT.

\subsection{Adding surface defects}

The above story can be further enriched by introduction of surface
defects.  Adding a surface defect that covers the holomorphic surface
$C$ modifies the chiral algebra of the theory.

As we did in section \ref{sec:CA-Omega}, let us think of the
topological--holomorphic twist as two successive two-dimensional
twists.  This time, however, we reverse the order and first perform
the twist along $\Sigma = \R^2$.  From the values of $M_\Sigma'$ for
the supercharges, we see that after this first twist we are left with
$\CN = (0,4)$ supersymmetry on $C$.  The second twist then gives a
twisted version of $\CN = (0,4)$ supersymmetry.

The supercharge $\QK$ relevant for the chiral algebra actually belongs
to the smaller subalgebra commuting with $M_\Sigma'$, generated by
$\QK$, $\CQ_-^2 + \CQt_{1\dot-}$, $\CP_{-\dot-}$, $M_C$ and $R$.  This
subalgebra is isomorphic to the $\CN = (0,2)$ supersymmetry algebra.
In the case of $\CN = 2$ SCFTs, it is further contained in the global
$\CN = (0,2)$ superconformal algebra, with the identification
\begin{equation}
  \begin{gathered}
    \Lb_0 = \frac12 \CH - \frac14 M_C \,,
    \qquad
    \Lb_{-1} = (\Lb_1)^\dagger = \CP_{-\dot-} \,,
    \qquad
    \Jb_0 = 2R \,,
    \\
    \Gb^+_{-\frac12}
    = (\Gb^-_{\frac12})^\dagger
    = \frac{1}{\sqrt{2}} \QK \,,
    \qquad
    \Gb^-_{-\frac12}
    = (\Gb^+_{\frac12})^\dagger
    = \frac{1}{\sqrt{2}} (\CQ^2_- + \CQt_{1\dot-}) \,.
    \qquad
  \end{gathered}
\end{equation}
Hence, we may take an $\CN = (0,2)$ supersymmetric field theory in two
dimensions, twist it, and couple it to the topological--holomorphic
theory along $\{0\} \times C \subset \R^2 \times C$.  From the
four-dimensional viewpoint, the two-dimensional theory creates a
surface defect at the origin of $\R^2$.  Note that in general it
preserves $M_\Sigma'$ but not $M_\Sigma$ and $r$ separately.  We let
$M_\Sigma'$ act trivially in the two-dimensional theory

The $\Omega$-deformation replaces $\QK$ with $\QK^\hbar$ and
$\CP_{-\dot-}$ with $\CP_{-\dot-} + 2\hbar \CR^2{}_1$, but together
with other generators they still generate a twisted $\CN = (0,2)$
supersymmetry subalgebra inside the two-dimensional theory where
$M_\Sigma' = 0$.  The $\QK^\hbar$-cohomology of local operators of the
twisted $\CN = (0,2)$ supersymmetric field theory is a chiral algebra
of its own.  Therefore, the $\QK^\hbar$-cohomology of local operators
of the coupled system is a chiral algebra constructed from those of
the four-dimensional theory and the two-dimensional theory.

As an example, take the gauge theory considered in
section~\ref{sec:VM-HM} as the four-dimensional theory, and couple to
it an $\CN = (0,2)$ chiral multiplet in a representation $R_C$ and a
Fermi multiplet in a representation $R_F$ of the gauge group $G$.  The
coupling is done via gauging.  The chiral algebras associated to the
$\CN = (0,2)$ multiplets are readily identified \cite{Witten:2005px,
  Tan:2006qt}.  The chiral multiplet yields the $\beta\gamma$ system
in the representation $R_C$.  In terms of the complex scalar $\phi$ of
the multiplet, the fields of this system are
\begin{equation}
  \betaSD = \del\phib \,,
  \qquad
  \gammaSD = \phi \,.
\end{equation}
The Fermi multiplet gives the $bc$ system in the representation $R_F$,
whose fields $\bSD$ and $\cSD$ are identified with the left-moving
fermions of the multiplet.  The chiral algebra of the coupled system
is therefore the BRST reduction of the $\beta\gamma$ system in the
representation $R_H \oplus R_C$ plus the $bc$ system in the
representation $\adj \oplus R_F$, governed by the action
\begin{equation}
  \frac{1}{\hbar} \int_C (\beta \delb\gamma + b \delb c
  + \betaSD \delb\gammaSD + \bSD \delb\cSD) \,,
\end{equation}
with the BRST charge given by
\begin{equation}
  Q_\BRST
  =
  \frac{1}{2\pi\iu\hbar} \oint \bigl(-\beta c \gamma + \Tr(bcc)
  - \betaSD c \gammaSD + \bSD c \cSD\bigr) \,.
\end{equation}

The surface defect has its own contribution to the anomaly in the BRST
symmetry.  The BRST charge of the coupled system satisfies the
relation
\begin{equation}
  \label{eq:BRST2-SD}
  Q_\BRST^2
  =
  \bigl(
  \Tr_{\adj \oplus R_F}(T_a T_b) - \Tr_{R_H \oplus R_C}(T_a T_b)
  \bigr) \del_w c^a c^b \,.
\end{equation}
Assuming that the four-dimensional theory has no $\U(1)_r$ anomaly,
the vanishing of $Q_\BRST^2$ requires
\begin{equation}
  \Tr_{R_C}(T_a T_b) - \Tr_{R_F}(T_a T_b) = 0 \,.
\end{equation}
This is the condition for the absence of gauge anomaly in the
two-dimensional theory.

While the absence of the BRST anomaly is indispensable for the
construction of the chiral algebra, $\U(1)_R$ does not have to be
anomaly free.  Even though $M_C'$ fails to be conserved if $\U(1)_R$
is anomalous, the chiral algebra can be defined without it.  This is
to be contrasted with $M_\Sigma'$ which makes an appearance in
$(\QK^\hbar)^2$.  Anomalies in $\U(1)_R$ may originate from
nonperturbative effects in the two-dimensional theory, and break
$\U(1)_R$ down to a finite subgroup.  Such a situation was studied
in~\cite{Witten:2005px, Tan:2008mi, Tan:2008mc, Yagi:2010tp} in the
context of $\CN = (0,2)$ supersymmetric sigma models, for which
$\U(1)_R$ anomalies are linked to conformal anomalies due to the
curvature of the target space.  It was shown in~\cite{Yagi:2010tp}
that the chiral algebras of models with $\U(1)_R$ anomalies lack an
energy--momentum tensor.  In extreme cases, the entire chiral algebra
vanish because worldsheet instantons make the identity operator
$Q$-exact~\cite{Tan:2008mi, Tan:2008mc, Yagi:2010tp}.

\subsection{Nonconformal theories with surface defects}

In view of the relation~\eqref{eq:BRST2-SD}, the condition
$Q_\BRST^2 = 0$ does not require the $\U(1)_r$ anomaly of the
four-dimensional theory and the gauge anomaly of the two-dimensional
theory to vanish separately.  All it asks is that the contributions to
$Q_\BRST^2$ from these anomalies cancel out.  This observation
suggests that it may be possible to define chiral algebras even for
theories with nonvanishing one-loop beta function if suitable surface
defects are inserted.%
\footnote{This possibility was suggested to us by Kevin Costello.  A
  three-dimensional analog of this mechanism has been considered
  in~\cite{Costello:2018fnz}.  A similar construction has also
  appeared in~\cite{DelZotto:2018tcj} in the context of BPS strings in
  six-dimensional SCFTs.}
As we now show, this is in fact true.  In particular, we can obtain
any nonanomalous gauged $\beta\gamma$--$bc$ system from an $\CN = 2$
supersymmetric gauge theory.

The key point is that $\U(1)_r$ anomalies by themselves are not
fundamental obstructions to the existence of the chiral algebra.
Rather, the real problem is that the twisted rotation symmetry on
$\Sigma = \R^2$ is anomalous, which means that $\QK^\hbar$ is not
conserved and the $\Omega$-deformation cannot be turned on.
\emph{Assuming} that the rotation symmetry on $\Sigma$ is unbroken,
$\U(1)_r$ anomalies imply that the twisted rotation symmetry is also
anomalous.  In principle, however, $M_\Sigma' = M_\Sigma + 2r$ may be
preserved while neither $M_\Sigma$ nor $r$ is.

We claim that this can be achieved by addition of the following term
to the action of the four-dimensional theory:
\begin{equation}
  -\frac{\iu}{4\pi^2}
  \int_{\R^2 \times C} \rmd \theta
  \bigl(\ChS_\adj(A) - \ChS_{R_H}(A)\bigr)
  \,,
\end{equation}
where
\begin{equation}
  \ChS_R(A)
  = \Tr_R\biggl(A \wedge \rmd A + \frac23 A \wedge A \wedge A\biggr)
\end{equation}
is the Chern--Simons three-form with the trace taken in the
representation $R$.  Formally, we can rewrite this term as
\begin{equation}
  \frac{\iu}{4\pi^2}
  \int_{\R^2 \times C} \theta
  \bigl(\Tr_\adj(F \wedge F) - \Tr_{R_H}(F \wedge F)\bigr) \,.
\end{equation}
This is much like the $\theta$-term, but here $\theta$ is not a
parameter: it is literally the angle coordinate on $\R^2$.  The second
expression makes it clear that although this term is multivalued, its
exponential is not.

Because of the explicit dependence on $\theta$, the above term breaks
the rotation symmetry on $\R^2$, unless the theory is superconformal
in which case this term simply vanishes.  The variation of the term
under the action of $e^{\iu \alpha M_\Sigma} \in \U(1)_\Sigma$ cancels
the anomalous phase factor~\eqref{eq:U(1)r-anomaly} of the fermionic
path integral measure, thereby restoring the twisted rotation symmetry
as desired.  We expect that the anomaly in $\QK^\hbar$ is also
canceled by the $\QK^\hbar$-variation of this term; although we are
not able to prove it here, this is very plausible given the
cancellation of the anomaly in $M_\Sigma'$.

The added term, however, is not gauge invariant.  Under the gauge
transformation $A \mapsto A + \rmd_A \chi$, it changes by
\begin{equation}
  -\frac{\iu}{4\pi^2}
  \int_{\R^2 \times C} \rmd\theta \,
  \rmd\bigl(\Tr_\adj(\chi \, \rmd A) - \Tr_{R_H}(\chi \, \rmd A)\bigr) \,.
\end{equation}
The integrand is not a total derivative since $\theta$ is not single
valued; it satisfies $\rmd^2\theta = 2\pi \delta_{0 \in \R^2}$, with
$\delta_{0 \in \R^2}$ being the delta two-form on $\R^2$ supported at
$0$.  Therefore, the gauge variation of the added term is localized at
the origin of $\R^2$:
\begin{equation}
  -\frac{\iu}{2\pi}
  \int_{\{0\} \times C}
  \bigl(\Tr_\adj(\chi \, \rmd A) - \Tr_{R_H}(\chi \, \rmd A)\bigr) \,.
\end{equation}
Since the integral is now performed over $\{0\} \times C$, it can be
canceled by an anomaly inflow~\cite{Callan:1984sa} from a
two-dimensional object that produces an equal and opposite gauge
anomaly.

The last integral is precisely of the form that the gauge anomalies of
two-dimensional chiral fermions take.  A left-moving fermion in a
representation $R$ induces the anomalous phase factor
\begin{equation}
  \exp\biggl(-\frac{\iu}{4\pi}
  \int_{\{0\} \times C} \Tr_R(\chi \, \rmd A)\biggr)
\end{equation}
in the path integral measure.  An $\CN = (0,2)$ chiral multiplet
contains a pair of right-moving fermions, while an $\CN = (0,2)$ Fermi
multiplet has a pair of left-moving fermions.  Thus, the gauge
variation is canceled by the gauge anomaly of a surface defect
constructed from a chiral multiplet valued in $R_C$ and a Fermi
multiplet valued in $R_F$ if and only if
\begin{equation}
  \exp\biggl(-\frac{\iu}{2\pi}
  \int_{\{0\} \times C}
  \bigl(
  \Tr_{\adj \oplus R_F}(T_a T_b) - \Tr_{R_H \oplus R_C}(T_a T_b)
  \bigr) (\chi^a \rmd A^b)\biggr)
  = 1
\end{equation}
for any gauge transformation parameter $\chi$.  This is equivalent to
the condition for $Q_\BRST$ to square to zero.

\section{\texorpdfstring{Topological quantum mechanics from
    three-dimensional $\CN = 4$ supersymmetric field
    theories}{Topological quantum mechanics from three-dimensional N =
    4 supersymmetric field theories}}
\label{sec:3d}

As dimensional reduction of the $\CN = 2$ superconformal algebra in
four dimensions yields the $\CN = 4$ superconformal algebra in three
dimensions, much of the analysis from the previous section carries
over to three-dimensional theories.  We briefly discuss how some known
results about $\CN = 4$ SCFTs can be understood in our approach.  We
follow the conventions of \cite{Beem:2016cbd}, except that we rescale
the supercharges by a factor of $1/2$.

The R-symmetry group of the $\CN = 4$ superconformal algebra is
$\SU(2)_H \times \SU(2)_C$.  In the dimensional reduction picture,
$\SU(2)_H$ comes from $\SU(2)_R$ in four dimensions, whereas the
diagonal subgroup of $\SU(2)_C$ can be identified with $\U(1)_r$.  We
denote the generators of $\SU(2)_H$ by $R^a{}_b$ and $\SU(2)_C$ by
$\Rt^{\at}{}_{\bt}$, and those of the rotation group $\SU(2)_M$ by
$M^\alpha{}_\beta$.  The eight Poincar\'e supercharges
$Q^{\alpha a\at}$ and the eight conformal supercharges
$S^{\alpha a\at}$ of the $\CN = 4$ superconformal algebra transform in
the trifundamental representation of
$\SU(2)_M \times \SU(2)_H \times \SU(2)_C$.  The remaining generators
are $P_{\alpha\beta}$ of translations, $D$ of dilatations and
$K^{\alpha\beta}$ of special conformal transformations.

The supercharges satisfy the commutation relations
\begin{equation}
  \begin{aligned}
    \{Q_\alpha^{a\at}, Q_\beta^{b\bt}\}
    &=
    \frac12 \eps^{ab} \eps^{\at\bt} P_{\alpha\beta} \,,
    \\
    \{S^\alpha_{a\at}, S^\beta_{b\bt}\}
    &=
    \frac12 \eps_{ab} \eps_{\at\bt} K^{\alpha\beta} \,,
    \\
    \{Q_\alpha^{a\at}, S^\beta_{b\bt}\}
    &=
    \frac12 \delta^a_b \delta^{\at}_{\bt}
    (M_\alpha{}^\beta + \delta_\alpha^\beta D)
    - \frac12 \delta_\alpha^\beta
    (R^a{}_b \delta^{\at}_{\bt} + \delta^a{}_b R^{\at}{}_{\bt}) \,,
  \end{aligned}
\end{equation}
and in radial quantization obey the hermiticity condition
\begin{equation}
  (Q_\alpha^{a\at})^\dagger = S^\alpha_{a\at} \,.
\end{equation}

To locate the $\CN = (2,2)$ superconformal algebra inside the
$\CN = 4$ superconformal algebra, we embed
$\U(1)_M \times \U(1)_V \times \U(1)_A$ into
$\SU(2)_M \times \SU(2)_H \times \SU(2)_C$.  As $\U(1)_V$ and
$\U(1)_A$ are identified with the diagonal subgroup of $\SU(2)_R$ and
$\U(1)_r$ in four dimensions, we take
\begin{equation}
  M = 2M^+{}_+ \,,
  \qquad
  F_V = 2R^1{}_1 \,,
  \qquad
  F_A = 2\Rt^{\tilde1}{}_{\tilde1} \,.
\end{equation}
Then we have
\begin{alignat}{4}
  \label{eq:Q-S-3d}
  Q_+ &= Q_+^{2\tilde2} \,,
  &\qquad
  Q_- &= Q_-^{2\tilde1} \,,
  &\qquad
  \Qb_+ &= Q_+^{1\tilde1} \,,
  &\qquad
  \Qb_- &= Q_-^{1\tilde2} \,,
  \\
  S_+ &= S^+_{1\tilde1} \,,
  &
  S_- &= S^-_{1\tilde2} \,,
  &
  \Sb_+ &= S^+_{2\tilde2} \,,
  &
  \Sb_- &= S^-_{2\tilde1} \,.
\end{alignat}
The commutation relations between the supercharges fix the
identification for the bosonic generators:
\begin{equation}
  \begin{alignedat}{2}
    L_{-1} &= -\frac12 P_{--} \,,
    &\qquad 
    \Lb_{-1} &= \frac12 P_{++} \,,   
    \\
    L_0 &= \frac12 (M^+{}_+ + D) \,,
    &
    \Lb_0 &= \frac12 (-M^+{}_+ + D) \,,
    \\
    L_1 &= -\frac12 K^{--} \,,
    &
    \Lb_1 &= \frac12 K^{++} \,,   
    \\
    J_0 &= R^1{}_1 - \Rt^{\tilde1}{}_{\tilde1} \,,
    &
    \Jb_0 &= R^1{}_1 + \Rt^{\tilde1}{}_{\tilde1} \,.
  \end{alignedat}
\end{equation}

Now, consider the topological--holomorphic twist of an $\CN = 2$
supersymmetric field theory on $\Sigma \times C$.  If we take
$C = \R \times S^1$ and perform dimensional reduction on $S^1$, we
obtain a topological twist of an $\CN = 4$ supersymmetric field theory
on $\Sigma \times \R$.  This sets the derivative along $S^1$ to zero,
hence turns the antiholomorphic derivative $\del_\wb$ on $C$ into the
derivative $\del_t$ along $\R$.  The relation $\del_\wb = 0$ in the
topological--holomorphic theory on $\Sigma \times C$ then implies that
we have $\del_t = 0$ after the reduction.  Therefore, the resulting
twisted theory on $\Sigma \times \R$ is fully topological.

There are two topological twists of a three-dimensional $\CN = 4$
supersymmetric field theory.  In the \emph{Rozansky--Witten
  twist}~\cite{Rozansky:1996bq}, the rotation group $\SU(2)_M$ is
replaced with the diagonal subgroup of $\SU(2)_M \times \SU(2)_C$ and
$Q$ is taken to be a linear combination of two supercharges that are
singlets under the twisted rotation group:
\begin{align}
  (M_\RW)^\alpha{}_\beta &= M^\alpha{}_\beta + \Rt^\alphat{}_\betat \,,
  \\
  Q_\RW &= Q_+^{1\tilde1} + Q_-^{1\tilde2} \,.
\end{align}
(Here we are using the indices $(+,-)$ and $(1,2)$ interchangeably.)
In the \emph{mirror Rozansky--Witten twist}~\cite{Kapustin:2010ag,
  Bullimore:2015lsa}, we use $\SU(2)_H$ to twist the rotation group:
\begin{align}
  (M_\mRW)^\alpha{}_\beta &= M^\alpha{}_\beta + R^\alpha{}_\beta \,,
  \\
  Q_\mRW &= Q_+^{1\tilde1} + Q_-^{2\tilde1} \,.
\end{align}
Under the identification~\eqref{eq:Q-S-3d} we have
\begin{align}
  Q_A &= Q_\mRW \,,
  \\
  Q_B &= Q_\RW \,.
\end{align}
Thus, from the point of view of the $\CN = (2,2)$ supersymmetry on
$\Sigma$, the Rozansky--Witten twist is the B-twist and the mirror
Rozansky--Witten twist is the A-twist.  The topological--holomorphic
twist reduces to the Rozansky--Witten twist.

The $Q$-cohomology of local operators in a topological twist of an
$\CN = 4$ supersymmetric field theory is called the \emph{chiral ring}
of the twisted theory.  It is necessarily commutative since there is
no invariant notion of operator ordering in three dimensions (or for
that matter, in any dimension greater than one).  For $\Sigma = \R^2$,
an $\Omega$-deformation introduces just such an ordering by forcing
local operators to line up on $\{0\} \times \R$.  Therefore, the
$Q^\hbar$-cohomology of local operators is a quantization of the
chiral ring.  Some aspects of the $\Omega$-deformed chiral rings of
$\CN = 4$ supersymmetric field theories were studied
in~\cite{Bullimore:2015lsa, Bullimore:2016nji, Bullimore:2016hdc,
  Costello:2016nkh, Costello:2017fbo, Dedushenko:2017avn,
  Beem:2018fng, Dedushenko:2018icp}.

For the Rozansky--Witten twist of an $\CN = 4$ supersymmetric gauge
theory constructed from vector multiplets and hypermultiplets, we can
easily identify the $\Omega$-deformed chiral ring.  The dimensional
reduction of the localization formula discussed in
section~\ref{sec:VM-HM} shows that the $\Omega$-deformed chiral ring
is the algebra of local operators in the $G_\C$-gauged quantum
mechanics, described by the action
\begin{equation}
  \frac{1}{\hbar} \int_\R (\qt \rmd_A q) \,.
\end{equation}
The Hamiltonian for this quantum mechanical system vanishes as a
consequence of the topological invariance on $\R$.  This result was
also obtained in~\cite{Dedushenko:2016jxl} via localization on~$S^3$.
The case of sigma models has been studied by localization
in~\cite{Yagi:2014toa}, and from the point of view of secondary
products and equivariant homology in~\cite{Beem:2018fng}.

For a unitary $\CN = 4$ SCFT on $\R^2 \times \R$, the
$Q_\RW^\hbar$-cohomology of local operators coincides with the
$\QQ_\RW^\hbar$-cohomology of local operators at the origin of $\R^2$.
According to the formula~\eqref{eq:QQB}, we have
\begin{equation}
  \QQ_\RW^\hbar = Q_+^{1\tilde1} + 2\hbar S^-_{1\tilde2} \,.
\end{equation}
This is the supercharge used in \cite{Beem:2016cbd} to define a
noncommutative operator algebra.  There, it was found that this
algebra is a deformation quantization of the chiral ring for the
Rozansky--Witten twist.  The above localization result, and the
results obtained in~\cite{Beem:2018fng}, explain why this is true.

\section*{Acknowledgments}

The authors are grateful to Kevin Costello for his advice and helpful
discussions. The research of JO is supported in part by Kwanjeong
Educational Foundation and by the Visiting Graduate Fellowship Program
at the Perimeter Institute for Theoretical Physics.  The research of
JY is supported by the Perimeter Institute for Theoretical Physics.
JY also acknowledges support of IH\'ES during his visit and funding from
the European Research Council (ERC) under the European Union's Horizon
2020 research and innovation program (QUASIFT grant agreement 677368).
Research at Perimeter Institute is supported by the Government of
Canada through Industry Canada and by the Province of Ontario through
the Ministry of Research and Innovation.

\appendix

\section{\texorpdfstring{$\Omega$-deformations for vector and chiral
    multiplets}{Ω-deformations for vector and chiral multiplets}}
\label{sec:QV-vector-chiral}

In this appendix, we describe $\Omega$-deformations of the topological
twists of $\CN = (2,2)$ supersymmetric gauge theories constructed from
vector multiplets and charged chiral multiplets.

In the A-twist, a vector multiplet consists of a gauge field $A$, a
complex scalar $\sigma$, an auxiliary scalar $\auxD$, fermionic
scalars $\alpha$, $\beta$ and a fermionic one-form $\lambda$:
\begin{equation}
  \sigma \,, \, \auxD \in \Omega^0 \,,
  \quad
  A \in \Omega^1 \,,
  \quad
  \alpha \,, \, \beta \in \Pi\Omega^0 \,,
  \quad
  \lambda \in \Pi\Omega^1 \,.
\end{equation}
Here $\Omega^p$ is the space of $p$-forms on the spacetime surface and
$\Pi$ denotes parity reversal.  A chiral multiplet (with $F_V = 0$)
consists of a complex scalar $\varphi$, an auxiliary one-form $\auxF$,
fermionic scalars $\eta$, $\etab$ and a fermionic one-form $\psi$:
\begin{equation}
  \varphi \in \Omega^0 \,,
  \quad
  \auxF \in \Omega^1 \,,
  \quad
  \eta \,, \, \etab \in \Pi\Omega^0 \,,
  \quad
  \psi \in \Pi\Omega^1 \,.
\end{equation}
The fields in the vector multiplet are valued in the adjoint
representation of the gauge group, whereas those in the chiral
multiplet are valued in some representation.

The $\Omega$-deformed supercharge $Q_V$ acts on the vector multiplet
by the transformation
\begin{equation}
  \begin{alignedat}{2}
    \delta_V A &= \lambda \,,
    &\qquad
    \delta_V\lambda &= \iota_V F - \rmd_A\sigma \,,
    \\
    \delta_V\sigma &= -\iota_V \lambda \,,
    &
    \delta_V\alpha &= \iota_V \rmd_A\sigmab + [\sigma,\sigmab]
    \\
    \delta_V \sigmab &= \alpha \,,
    &
    \delta_V\beta &= \auxD \,,
    \\
    \delta_V \auxD &= \iota_V \rmd_A\beta + [\sigma,\beta] \,,
  \end{alignedat}
\end{equation}
where $F = \rmd A + A \wedge A$ is the curvature of $A$.  On the
chiral multiplet, $Q_V$ acts by
\begin{equation}
  \begin{alignedat}{2}
    \delta_V\varphi &= \eta \,,
    &
    \delta_V\varphib &= \etab \,,
    \\
    \delta_V\eta &= \iota_V\rmd_A\varphi + \sigma\phi \,,
    &\qquad
    \delta_V\etab &= \iota_V\rmd_A\phib + \sigma\phib \,,
    \\
    \delta_V\psi &= \auxF \,,
    &
    \delta_V\auxF &= (\rmd_A \iota_V + \iota_V \rmd_A) \psi + \sigma\psi \,.
  \end{alignedat}
\end{equation}
The square of this transformation is the gauge covariant Lie
derivative with respect to $V$ plus the infinitesimal gauge
transformation generated by $\sigma$:
\begin{equation}
  \delta_V^2 = \rmd_A \iota_V + \iota_V \rmd_A + \delta_\sigma \,.
\end{equation}

For the B-twist, a vector multiplet contains a one-form $\sigma$
rather than a complex scalar, and a two-form fermion $\zeta$ instead
of a scalar fermion:
\begin{equation}
  \auxD \in \Omega^0 \,,
  \quad
  A \,, \, \sigma \in \Omega^1 \,,
  \quad
  \alpha \in \Pi\Omega^0 \,,
  \quad
  \lambda \in \Pi\Omega^1 \,,
  \quad
  \zeta \in \Pi\Omega^2 \,,
\end{equation}
A chiral multiplet (with $F_A = 0$) consists of
\begin{equation}
  \varphi \in \Omega^0 \,,
  \quad
  \auxF \in \Omega^2 \,,
  \quad
  \etab \in \Pi\Omega^0 \,,
  \quad
  \rho \in \Pi\Omega^1 \,,
  \quad
  \mub \in \Pi\Omega^2 \,,
\end{equation}
where $\auxF$ is a complex two-form.

The action of $Q_V$ is given by
\begin{equation}
  \begin{alignedat}{2}
    \delta_V\CA &= \iota_V\zeta \,,
    &
    \delta_V\CAb &= \lambda - \iota_V\zeta \,,
    \\
    \delta_V\lambda &= 2\iota_V F - 2\iu  D\iota_V\sigma \,,
    &\qquad
    \delta_V\zeta &= \CF \,,
    \\
    \delta_V\alpha &= \auxD \,,
    &
    \delta_V \auxD &= \iota_V\cD\alpha
  \end{alignedat}
\end{equation}
for the vector multiplet and
\begin{equation}
  \begin{alignedat}{2}
    \delta_V\varphi &= \iota_V\rho \,,
    &
    \delta_V\varphib &= \etab \,,
    \\
    \delta_V\rho &= \cD\varphi + \iota_V\auxF \,,
    &\qquad
    \delta_V\etab &= \iota_V \cD \varphib \,,
    \\
    \delta_V\auxF &= \cD\rho - \zeta\varphi \,,
    &
    \delta_V\auxFb &= \cD\iota_V \mub \,,
    \\
    &&
    \delta_V\mub &= \auxFb \,.
  \end{alignedat}
\end{equation}
for the chiral multiplet.  Its square is the covariant Lie derivative
with respect to the complexified gauge field $\CA = A + \iu\sigma$:
\begin{equation}
  \delta_V^2
  = \rmd_\CA \iota_V + \iota_V \rmd_\CA
  \,.
\end{equation}

\bibliographystyle{../JHEP}
\bibliography{../junya}

\providecommand{\href}[2]{#2}\begingroup\raggedright\begin{thebibliography}{10}

\bibitem{Beem:2013sza}
C.~Beem, M.~Lemos, P.~Liendo, W.~Peelaers, L.~Rastelli and B.~C. van Rees,
  \emph{Infinite chiral symmetry in four dimensions},
  \href{http://dx.doi.org/10.1007/s00220-014-2272-x}{\emph{Comm. Math. Phys.}
  {\bf 336} (2015) 1359--1433} [\href{http://arxiv.org/abs/1312.5344}{{\tt
  1312.5344}}].

\bibitem{Liendo:2015ofa}
P.~Liendo, I.~Ram\'{\i}rez and J.~Seo, \emph{Stress--tensor {OPE} in
  {$\mathcal{N}=2$} superconformal theories},
  \href{http://dx.doi.org/10.1007/JHEP02(2016)019}{\emph{JHEP} \textbf{02} (2016) 019} [\href{http://arxiv.org/abs/1509.00033}{{\tt 1509.00033}}].

\bibitem{Lemos:2015orc}
M.~Lemos and P.~Liendo, \emph{{$\mathcal{N}=2$} central charge bounds from
  {$2d$} chiral algebras},
  \href{http://dx.doi.org/10.1007/JHEP04(2016)004}{\emph{JHEP} \textbf{04} (2016) 004} [\href{http://arxiv.org/abs/1511.07449}{{\tt 1511.07449}}].

\bibitem{Kapustin:2006hi}
A.~Kapustin, \emph{{Holomorphic reduction of $\mathcal{N} = 2$ gauge theories,
  Wilson--'t Hooft operators, and S-duality}},
  \href{http://arxiv.org/abs/hep-th/0612119}{{\tt hep-th/0612119}}.

\bibitem{Witten:1988ze}
E.~Witten, \emph{Topological quantum field theory}, {\emph{Comm. Math. Phys.}
  {\bf 117} (1988) 353--386}.

\bibitem{Witten:1988xj}
E.~Witten, \emph{Topological sigma models}, {\emph{Comm. Math. Phys.} {\bf 118}
  (1988) 411--449}.

\bibitem{Costello:2018fnz}
K.~Costello and D.~Gaiotto, \emph{{Vertex Operator Algebras and 3d $\mathcal
  N=4$ gauge theories}},  \href{http://arxiv.org/abs/1804.06460}{{\tt
  1804.06460}}.

\bibitem{Nekrasov:2002qd}
N.~A. Nekrasov, \emph{Seiberg--{W}itten prepotential from instanton counting},
  {\emph{Adv. Theor. Math. Phys.} {\bf 7} (2003) 831--864}
  [\href{http://arxiv.org/abs/hep-th/0206161}{{\tt hep-th/0206161}}].

\bibitem{Nekrasov:2003rj}
N.~A. Nekrasov and A.~Okounkov, \emph{Seiberg--{W}itten theory and random
  partitions},  in \emph{The unity of mathematics}, vol.~244 of \emph{Progr.
  Math.}, p.~525.
\newblock Birkh{\"a}user Boston, Boston, MA, 2006.
\newblock \href{http://arxiv.org/abs/hep-th/0306238}{{\tt hep-th/0306238}}.
\newblock \href{http://dx.doi.org/10.1007/0-8176-4467-9_15}{DOI}.

\bibitem{Butson:2019}
D.~Butson, \emph{Omega backgrounds and boundary theories in twisted
  supersymmetric gauge theories}, in preparation.

\bibitem{Rozansky:1996bq}
L.~Rozansky and E.~Witten, \emph{Hyper-{K}{\"a}hler geometry and invariants of
  three-manifolds},
  \href{http://dx.doi.org/10.1007/s000290050016}{\emph{Selecta Math. (N.S.)}
  {\bf 3} (1997) 401} [\href{http://arxiv.org/abs/hep-th/9612216}{{\tt
  hep-th/9612216}}].

\bibitem{Beem:2016cbd}
C.~Beem, W.~Peelaers and L.~Rastelli, \emph{Deformation quantization and
  superconformal symmetry in three dimensions},
  \href{http://dx.doi.org/10.1007/s00220-017-2845-6}{\emph{Comm. Math. Phys.}
  {\bf 354} (2017) 345--392} [\href{http://arxiv.org/abs/1601.05378}{{\tt
  1601.05378}}].

\bibitem{Dedushenko:2016jxl}
M.~Dedushenko, S.~S. Pufu and R.~Yacoby, \emph{A one-dimensional theory for
  {H}iggs branch operators},
  \href{http://dx.doi.org/10.1007/JHEP03(2018)138}{\emph{JHEP} \textbf{03} (2018) 138} [\href{http://arxiv.org/abs/1610.00740}{{\tt 1610.00740}}].

\bibitem{Vafa:1990mu}
C.~Vafa, \emph{Topological {L}andau--{G}inzburg models},
  \href{http://dx.doi.org/10.1142/S0217732391000324}{\emph{Mod. Phys. Lett.
  A} {\bf 6} (1991) 337}.

\bibitem{Witten:1991zz}
E.~Witten, \emph{Mirror manifolds and topological field theory},  in
  \emph{Essays on mirror manifolds}, pp.~120--158.
\newblock Int. Press, Hong Kong, 1992.
\newblock \href{http://arxiv.org/abs/hep-th/9112056}{{\tt hep-th/9112056}}.

\bibitem{Yagi:2014toa}
J.~Yagi, \emph{{$\Omega$}-deformation and quantization},
  \href{http://dx.doi.org/10.1007/JHEP08(2014)112}{\emph{JHEP} \textbf{08} (2014) 112} [\href{http://arxiv.org/abs/1405.6714}{{\tt 1405.6714}}].

\bibitem{Luo:2014sva}
Y.~Luo, M.-C. Tan, J.~Yagi and Q.~Zhao, \emph{{$\Omega$}-deformation of
  {B}-twisted gauge theories and the 3d-3d correspondence},
  \href{http://dx.doi.org/10.1007/JHEP02(2015)047}{\emph{JHEP} \textbf{02} (2015) 047} [\href{http://arxiv.org/abs/1410.1538}{{\tt 1410.1538}}].

\bibitem{Nekrasov:2018pqq}
N.~Nekrasov, \emph{{Tying up instantons with anti-instantons}},
  \href{http://arxiv.org/abs/1802.04202}{{\tt 1802.04202}}.

\bibitem{Costello:2018txb}
K.~Costello and J.~Yagi, \emph{Unification of integrability in supersymmetric
  gauge theories},  \href{http://arxiv.org/abs/1810.01970}{{\tt 1810.01970}}.

\bibitem{Cordova:2017mhb}
C.~C\'{o}rdova, D.~Gaiotto and S.-H. Shao, \emph{Surface defects and chiral
  algebras}, \href{http://dx.doi.org/10.1007/JHEP05(2017)140}{\emph{JHEP} \textbf{05}
  (2017) 140} [\href{http://arxiv.org/abs/1704.01955}{{\tt
  1704.01955}}].

\bibitem{Pan:2017zie}
Y.~Pan and W.~Peelaers, \emph{Chiral algebras, localization and surface
  defects}, \href{http://dx.doi.org/10.1007/jhep02(2018)138}{\emph{JHEP} \textbf{02} (2018)
  138} [\href{http://arxiv.org/abs/1710.04306}{{\tt
  1710.04306}}].

\bibitem{Friedan:1985ge}
D.~Friedan, E.~Martinec and S.~Shenker, \emph{Conformal invariance,
  supersymmetry and string theory},
  \href{http://dx.doi.org/10.1016/0550-3213(86)90356-1}{\emph{Nucl. Phys. B}
  {\bf 271} (1986) 93--165}.

\bibitem{Pan:2019bor}
Y.~Pan and W.~Peelaers, \emph{{Schur correlation functions on $S^3\times
  S^1$}},  \href{http://arxiv.org/abs/1903.03623}{{\tt 1903.03623}}.

\bibitem{Dedushenko:2019yiw}
M.~Dedushenko and M.~Fluder, \emph{Chiral algebra, localization, modularity,
  surface defects, and all that},  \href{http://arxiv.org/abs/1904.02704}{{\tt
  1904.02704}}.

\bibitem{Witten:2005px}
E.~Witten, \emph{Two-dimensional models with {$(0,2)$} supersymmetry:
  perturbative aspects}, {\emph{Adv. Theor. Math. Phys.} {\bf 11} (2007) 1}
  [\href{http://arxiv.org/abs/hep-th/0504078}{{\tt hep-th/0504078}}].

\bibitem{Tan:2006qt}
M.-C. Tan, \emph{Two-dimensional twisted sigma models and the theory of chiral
  differential operators}, {\emph{Adv. Theor. Math. Phys.} {\bf 10} (2006)
  759--851} [\href{http://arxiv.org/abs/hep-th/0604179}{{\tt
  hep-th/0604179}}].

\bibitem{Tan:2008mi}
M.-C. Tan and J.~Yagi, \emph{Chiral algebras of $(0,2)$ sigma models: beyond
  perturbation theory},
  \href{http://dx.doi.org/10.1007/s11005-008-0249-4}{\emph{Lett. Math. Phys.}
  {\bf 84} (2008) 257} [\href{http://arxiv.org/abs/0801.4782}{{\tt
  0801.4782}}].

\bibitem{Tan:2008mc}
M.-C. Tan and J.~Yagi, \emph{Chiral algebras of $(0,2)$ sigma models: beyond
  perturbation theory -- {II}},
  \href{http://dx.doi.org/10.1007/s11005-008-0249-4}{\emph{Lett. Math. Phys.}
  {\bf 84} (2008) 257} [\href{http://arxiv.org/abs/0805.1410}{{\tt
  0805.1410}}].

\bibitem{Yagi:2010tp}
J.~Yagi, \emph{Chiral algebras of {$(0,2)$} models}, {\emph{Adv. Theor. Math.
  Phys.} {\bf 16} (2012) 1} [\href{http://arxiv.org/abs/1001.0118}{{\tt
  1001.0118}}].

\bibitem{DelZotto:2018tcj}
M.~Del~Zotto and G.~Lockhart, \emph{Universal features of {BPS} strings in
  six-dimensional {SCFT}s},
  \href{http://dx.doi.org/10.1007/jhep08(2018)173}{\emph{JHEP} \textbf{08} (2018) 173} [\href{http://arxiv.org/abs/1804.09694}{{\tt
  1804.09694}}].

\bibitem{Callan:1984sa}
C.~G. Callan, Jr. and J.~A. Harvey, \emph{Anomalies and fermion zero modes on
  strings and domain walls},
  \href{http://dx.doi.org/10.1016/0550-3213(85)90489-4}{\emph{Nucl. Phys. B}
  {\bf 250} (1985) 427--436}.

\bibitem{Kapustin:2010ag}
A.~Kapustin and K.~Vyas, \emph{A-models in three and four dimensions},
  \href{http://arxiv.org/abs/1002.4241}{{\tt 1002.4241}}.

\bibitem{Bullimore:2015lsa}
M.~Bullimore, T.~Dimofte and D.~Gaiotto, \emph{The {C}oulomb branch of 3d
  {$\mathcal{N}=4$} theories},
  \href{http://dx.doi.org/10.1007/s00220-017-2903-0}{\emph{Comm. Math. Phys.}
  {\bf 354} (2017) 671--751} [\href{http://arxiv.org/abs/1503.04817}{{\tt
  1503.04817}}].

\bibitem{Bullimore:2016nji}
M.~Bullimore, T.~Dimofte, D.~Gaiotto and J.~Hilburn, \emph{Boundaries, mirror
  symmetry, and symplectic duality in 3d $\mathcal{N}=4$ gauge theory},
  \href{http://dx.doi.org/10.1007/JHEP10(2016)108}{\emph{JHEP} {\bf 10} (2016)
  108} [\href{http://arxiv.org/abs/1603.08382}{{\tt 1603.08382}}].

\bibitem{Bullimore:2016hdc}
M.~Bullimore, T.~Dimofte, D.~Gaiotto, J.~Hilburn and H.-C. Kim, \emph{Vortices
  and {V}ermas},  \href{http://arxiv.org/abs/1609.04406}{{\tt 1609.04406}}.

\bibitem{Costello:2016nkh}
K.~Costello, \emph{{M}-theory in the {$\Omega$}-background and 5-dimensional
  non-commutative gauge theory},  \href{http://arxiv.org/abs/1610.04144}{{\tt
  1610.04144}}.

\bibitem{Costello:2017fbo}
K.~Costello, \emph{Holography and {K}oszul duality: the example of the {$M2$}
  brane},  \href{http://arxiv.org/abs/1705.02500}{{\tt 1705.02500}}.

\bibitem{Dedushenko:2017avn}
M.~Dedushenko, Y.~Fan, S.~S. Pufu and R.~Yacoby, \emph{Coulomb branch operators
  and mirror symmetry in three dimensions},
  \href{http://dx.doi.org/10.1007/JHEP04(2018)037}{\emph{JHEP} \textbf{04} (2018) 037} [\href{http://arxiv.org/abs/1712.09384}{{\tt
  1712.09384}}].

\bibitem{Beem:2018fng}
C.~Beem, D.~Ben-Zvi, M.~Bullimore, T.~Dimofte and A.~Neitzke, \emph{{Secondary
  products in supersymmetric field theory}},
  \href{http://arxiv.org/abs/1809.00009}{{\tt 1809.00009}}.

\bibitem{Dedushenko:2018icp}
M.~Dedushenko, Y.~Fan, S.~S. Pufu and R.~Yacoby, \emph{Coulomb branch
  quantization and abelianized monopole bubbling},
  \href{http://arxiv.org/abs/1812.08788}{{\tt 1812.08788}}.

\end{thebibliography}\endgroup

\end{document}